\documentclass{JFM-FLM_Au}
\usepackage{float}
\usepackage{placeins}

\lefttitle{S. Inparaja, C. Vreugdenhil and B. Gayen}
\righttitle{Journal of Fluid Mechanics}

\title{Buoyancy-driven circulation and topographic Rossby waves over an idealised Antarctic continental slope}

\author{Sivakunalan Inparaja\aff{1,2}, Catherine Vreugdenhil\aff{1,2} \and Bishakhdatta Gayen\aff{1,2,3}}

\affiliation{\aff{1}Department of Mechanical Engineering, The University of Melbourne, Melbourne, VIC, Australia
\aff{2}The Australian Centre for Excellence in Antarctic Science, The University of Melbourne, Melbourne, VIC, Australia
\aff{3}The Centre for Atmospheric and Oceanic Sciences, Indian Institute of Science, India}

\corresau{Sivakunalan Inparaja, \email{sivakunalani@student.unimelb.edu.au}}

\begin{document}
\maketitle
\begin{abstract}
Direct numerical simulations are used to investigate surface buoyancy-driven circulation over an idealised continental shelf--slope system. The configuration is motivated by the Antarctic continental margin and is used to study the role of rotation and topography in a single-active-scalar model. A non-rotating case and three rotating cases spanning Rossby numbers $Ro=1.4\times10^{-1}$ to $8.9\times10^{-3}$ are examined at a horizontal Rayleigh number of $Ra=3.24\times10^{12}$ and a Prandtl number of $Pr=5$. Without rotation, surface thermal forcing produces a broad downslope flow and domain-scale overturning. Rotation reorganises the circulation into oppositely directed slope-attached and surface-intensified along-slope currents and strongly suppresses interior cross-slope and vertical motions. Both along-slope transport branches weaken as rotation increases, with the surface-intensified branch showing a stronger Rossby-number dependence than the slope-attached branch. Residual overturning shows distinct interior and near-surface cells, with the interior cell weakening more strongly as rotation increases. The upper thermal boundary layer thickens as $\delta_T/L\propto Ro^{-0.21}$, whereas the bottom Ekman boundary layer thins as $\delta_E/L\propto Ro^{0.25}$, with bottom-layer transport generally dominated by the mean flow. Temporal fluctuations account for most of the full-column cross-slope transport of dense water, while the bottom boundary layer contributes less than $5\,\%$ of the downslope dense-water transport. Energy spectral analysis reveals a low-frequency ridge that is qualitatively consistent with the linear topographic Rossby-wave dispersion relation. The phase speed varies weakly with rotation over the upper slope but decreases with stronger rotation over the lower slope. These results reveal mean near-bottom transport, substantial fluctuating cross-slope exchange, and low-frequency variability consistent with topographic Rossby waves.
\end{abstract}

\section{Introduction}
\label{sec:introduction}
In polar oceans, dense water is formed through surface buoyancy loss associated with atmospheric cooling and brine rejection during sea-ice production. Around Antarctica, these processes occur predominantly over the continental shelf, where cold, saline water sinks, spreads towards the shelf break, and is subsequently exported down the continental slope. During its descent, it mixes with the surrounding fluid and undergoes dynamical processes that determine its transport and evolution. The resulting dense water contributes to Antarctic Bottom Water, which ventilates the abyssal ocean and spreads towards lower latitudes \citep{killworth1977mixing,baines1998observations}. Although the overall pathway of dense-water formation and export is well established, how dynamical processes over the continental slope regulate overturning and the along- and cross-slope transport of dense water remains less well understood. 

Idealised laboratory and numerical studies have provided much of the understanding of rotating dense-current transport over uniform slopes. In such configurations, dense water initially tends to accelerate downslope under gravity, while rotation can promote geostrophic adjustment and the development of a predominantly along-slope current \citep{smith1977experiments,lane1998eddy}. Bottom friction enables cross-slope transport through a slope-adjacent Ekman layer, allowing continued downslope drainage beneath the geostrophic flow \citep{smith1977experiments,nagata1993laboratory,condie1995descent}. The characteristic thickness of this boundary layer is expected to scale with the classical Ekman layer thickness, $\delta_E\sim(2\nu/f)^{1/2}$, where $\nu$ is the kinematic viscosity and $f$ is the Coriolis parameter. \citet{condie1995descent} also showed that, when dense-water production exceeds the downslope transport that can be accommodated within the bottom boundary layer, dense fluid accumulates, causing the lateral density front to steepen and the associated along-slope current to strengthen. In addition to cross-slope transport through the bottom Ekman layer, baroclinic instability of the dense current can generate eddies that provide an additional mechanism for transporting dense water across the slope. Numerical simulations by \citet{tanaka2006effects} also showed that downslope transport during the initial development of the current is broadly consistent with bottom Ekman-layer theory, whereas after baroclinic instability reaches finite amplitude, eddy transport becomes the primary mechanism for the offshore transport of dense water, although bottom Ekman transport can remain dynamically significant.

The unsteady behaviour of rotating dense currents is sensitive to the forcing and parameter regime. Small perturbations of the density front or of the interface between the dense current and the overlying ambient fluid can amplify into wave-like disturbances and finite-amplitude frontal deformations \citep{Griffiths_Killworth_Stern_1982,choboter2000baroclinic}. In some configurations, instability of the dense current produces eddies associated with deformation of the density front. A distinct response can arise through dynamical coupling between the dense lower layer and the overlying ambient fluid, producing vortical motions that remain coherent through a substantial fraction of the water column \citep{mory1987coherent,whitehead1990experimental,lane1998eddy}. Such vertically coherent vortical structures may subsequently propagate along isobaths \citep{nof1983translation}. Laboratory experiments have further identified laminar, wave-dominated and eddy-dominated regimes whose occurrence depends on parameters including the rotation rate, dense-water supply, density contrast and bottom-slope angle \citep{ADenseCurrentFlowingdownaSlopingBottominaRotatingFluid,sutherland2004rotating}. Rotating dense currents can therefore develop a range of unsteady responses, from frontal waves and eddies to vertically coherent vortical motions, with the resulting flow regime depending on the forcing and governing parameters.

The studies discussed above demonstrate that geostrophic adjustment and the resulting along-slope currents, bottom Ekman transport, and flow instability play important roles in controlling the evolution and cross-slope transport of rotating dense currents. In many of these studies, however, dense water is introduced through a prescribed source or generated by buoyancy forcing at a lateral boundary, such that the primary focus is on the subsequent evolution of an established dense current over sloping topography. In polar shelf seas, by contrast, dense water can be generated locally by surface buoyancy loss, so that dense-water formation, the resulting overturning circulation, and its subsequent export across the shelf and slope form a dynamically coupled system. Surface-forced convection over sloping topography has been examined in non-rotating configurations. For example, \citet{ulloa2022development} considered a numerical simulation of uniform surface heat loss over a sloping domain and showed that shallower waters cool more rapidly than deeper waters, generating horizontal density and pressure gradients that drive a downslope dense current. They further derived characteristic time and velocity scales associated with the transition from Rayleigh--Bénard-type convection to a large-scale overturning circulation over the shallower sloping section. Here, the motivation is to understand how this coupled pathway of dense-water formation, overturning, and cross-slope export is modified when surface buoyancy forcing acts together with rotation and continental shelf--slope topography.

A natural framework for investigating such buoyancy-driven large-scale circulations is horizontal convection, in which differential buoyancy forcing is applied along a horizontal boundary, generating lateral density gradients that drive an overturning circulation. Originally introduced by \citet{rossby1965thermal} as an idealised representation of large-scale oceanic buoyancy contrasts, horizontal convection has subsequently been used to investigate overturning circulation, boundary-layer dynamics, turbulent mixing, deep-ocean ventilation and energetics \citep{mullarney2004convection, annurev:/content/journals/10.1146/annurev.fluid.40.111406.102148, griffiths2013horizontal,stewart2012mixing,gayen2013energetics,GayenKlocker2024}. Rotation promotes geostrophic balance in the interior, reorganises the overturning circulation, and modifies the thermal boundary layer and associated heat transport. In rotating horizontal convection, scaling arguments based on geostrophic balance, continuity, and thermal transport have also been used to relate the upper thermal-boundary-layer structure to the strength of rotation \citep{Hignett_Ibbetson_Killworth_1981,barkan2013rotating, vreugdenhil2017geostrophic, sohail2018convection, Vreugdenhil_Gayen_Griffiths_2019, gayen2022rotating}. Previous studies of rotating horizontal convection, however, have focused predominantly on flat-bottom geometries. The combined influence of surface dense-water formation, rotation, and continental shelf--slope topography on the resulting flow dynamics therefore remains comparatively unexplored.

Sloping topography can support low-frequency topographic Rossby waves. Cross-slope motion over variations in water depth stretches or compresses fluid columns, producing a potential-vorticity restoring mechanism that supports predominantly along-slope wave propagation \citep{pedlosky1967laboratory,rhines1970edge,pedlosky1987}. In a stratified fluid, the wave properties depend on rotation, stratification, topographic slope, water-column depth, horizontal wavenumber, and the background flow. Connections between dense-water outflows and topographically controlled variability have been identified in previous numerical studies. 

A large-scale idealised numerical study by \citet{marques2014topographic} demonstrated that Antarctic dense-shelf-water outflows can excite topographic vorticity waves (TVWs), whose properties depend on the bottom slope, density anomaly, and background along-slope flow. Another large-scale idealised numerical study by \citet{han2022trw} showed that baroclinic instability of dense overflows can generate topographic Rossby waves, which can subsequently feed back on the overflow and produce coherent subsurface eddies. More recently, \citet{han2023controls} investigated the relationship between topographic Rossby-wave properties and downslope dense-water transport. They identified distinct regimes in which bottom friction dominates downslope transport over steep slopes, wave-associated interfacial form stress drives relatively rapid downslope transport over intermediate slopes, and coherent eddies inhibit downslope transport over gentle slopes.

These large-scale idealised numerical studies demonstrate that topographically controlled waves and eddies can play important roles in the dynamics of dense overflows. In their configurations, dense shelf water is supplied through a prescribed inflow, while an ambient along-slope current may be imposed separately. Important processes controlling dense-overflow transport and mixing occur at scales that are not fully resolved in large-scale ocean models and therefore often require parameterisation \citep{legg2009improving}. A clear understanding of these slope dynamics is therefore important for improving the representation of dense overflows in such models.

In contrast to the prescribed-overflow configurations, the present study considers continuously surface-forced horizontal convection in which the density structure, overturning circulation, and along-slope currents develop in response to the imposed surface temperature distribution. This configuration allows the effects of rotation on the mean circulation, slope-attached boundary-layer transport, and low-frequency wave variability to be examined within the same buoyancy-driven system. However, how increasing rotation modifies the characteristics of topographic waves in such a surface-forced shelf--slope system remains comparatively unexplored.

Idealised Direct numerical simulation at laboratory scales provides a controlled setting in which the fundamental dynamics of horizontal convection can be examined systematically. Using such configurations, turbulence-resolving simulations have investigated overturning circulation, boundary-layer dynamics, and energetics over computationally accessible ranges of the governing parameters \citep{gayen2013energetics,gayen2014stability,Vreugdenhil_Gayen_Griffiths_2019}. Because direct numerical simulations cannot reproduce the full range of oceanic length scales and extremely large Rayleigh numbers, estimated to be $Ra=1.2\times10^{25}$--$1.9\times10^{26}$ for the Southern Ocean \citep{sohail2018convection}, these simulations are primarily used to identify fundamental physical mechanisms and nondimensional trends rather than reproduce oceanic circulation at its actual scale. Their relevance is therefore assessed through the governing nondimensional parameters.

The present study adopts this approach using direct numerical simulations of surface-forced rotating horizontal convection over an idealised continental shelf--slope system at reduced, computationally accessible laboratory scales. The model geometry is motivated by the Antarctic continental shelf--slope structure and draws on the idealised large-scale ocean model configurations of \citet{stewart2013connecting,stewart2015eddy,stewart2016eddy}. A non-rotating reference case and three rotating cases spanning Rossby numbers from $Ro=1.4\times10^{-1}$ to $8.9\times10^{-3}$ are considered, while the remaining control parameters are held fixed. We quantify how rotation reorganises the buoyancy-driven circulation and transport, focusing on the along-slope and cross-slope flows, residual overturning circulation, upper thermal boundary layer, and slope-attached bottom boundary layer. We then analyse the spontaneously generated low-frequency slope variability using frequency--wavenumber spectra and compare its dispersion characteristics with linear topographic Rossby-wave theory, thereby testing whether classical topographic Rossby-wave dynamics emerge within the system. The remainder of the paper is organised as follows. Section~\ref{sec:numerical_methodology} describes the numerical configuration, governing equations, control parameters, and numerical method. Section~\ref{sec:Results} examines the rotational response of the mean circulation and boundary layers, followed by the low-frequency slope variability and its relationship to topographic Rossby-wave dynamics. Section~\ref{sec:discussion} discusses the interpretation of the results and their relevance to Antarctic continental-slope flows, and section~\ref{sec:conclusions} summaries the conclusions.

\section{Numerical methodology}
\label{sec:numerical_methodology}

We consider a three-dimensional rotating fluid domain representing an idealised continental shelf--slope system, as shown in figure~\ref{fig:model}. A Cartesian coordinate system $(x,y,z)$ is adopted, where $x$, $y$ and $z$ denote the along-slope (zonal), cross-slope (meridional) and vertical directions, respectively, with $z$ increasing upwards. The cross-slope domain width is taken as the characteristic length scale, $L\equiv L_y$, and all spatial coordinates are nondimensionalised according to $\hat{x}=x/L$, $\hat{y}=y/L$ and $\hat{z}=z/L$. The computational domain therefore extends over $0\leq\hat{x}\leq L_x/L=4/3$, $0\leq\hat{y}\leq1$ and $\hat{z}_b(\hat{y})\leq\hat{z}\leq H/L=0.4$, where $\hat{z}_b=z_b/L$ denotes the dimensionless bottom elevation. The bottom topography varies only in the cross-slope direction and is prescribed using a hyperbolic-tangent profile,

\begin{equation}
\hat{z}_b(\hat{y})
=
\frac{\hat{H}_s}{2}
\left[
1-\tanh\left(
\frac{0.6(\hat{y}-\hat{Y}_s)}{\hat{W}_s}
\right)
\right],
\label{eq:bathymetry}
\end{equation}
where $\hat{H}_s=1/3$ is the dimensionless shelf elevation, $\hat{Y}_s=1/3$ specifies the centre of the shelf--slope transition, and $\hat{W}_s=1/15$ controls the width of the slope region. The fluid is subject to gravitational acceleration $\boldsymbol{g}=-g\boldsymbol{e}_z$, where $\boldsymbol{e}_z$ is the unit vector in the positive vertical direction. The system rotates uniformly about the vertical axis with angular velocity $\boldsymbol{\Omega}$, and the corresponding Coriolis parameter is $f=2\boldsymbol{\Omega}\cdot\boldsymbol{e}_z$.

\begin{figure}[h]
\centering
\includegraphics[width=1.0\textwidth]{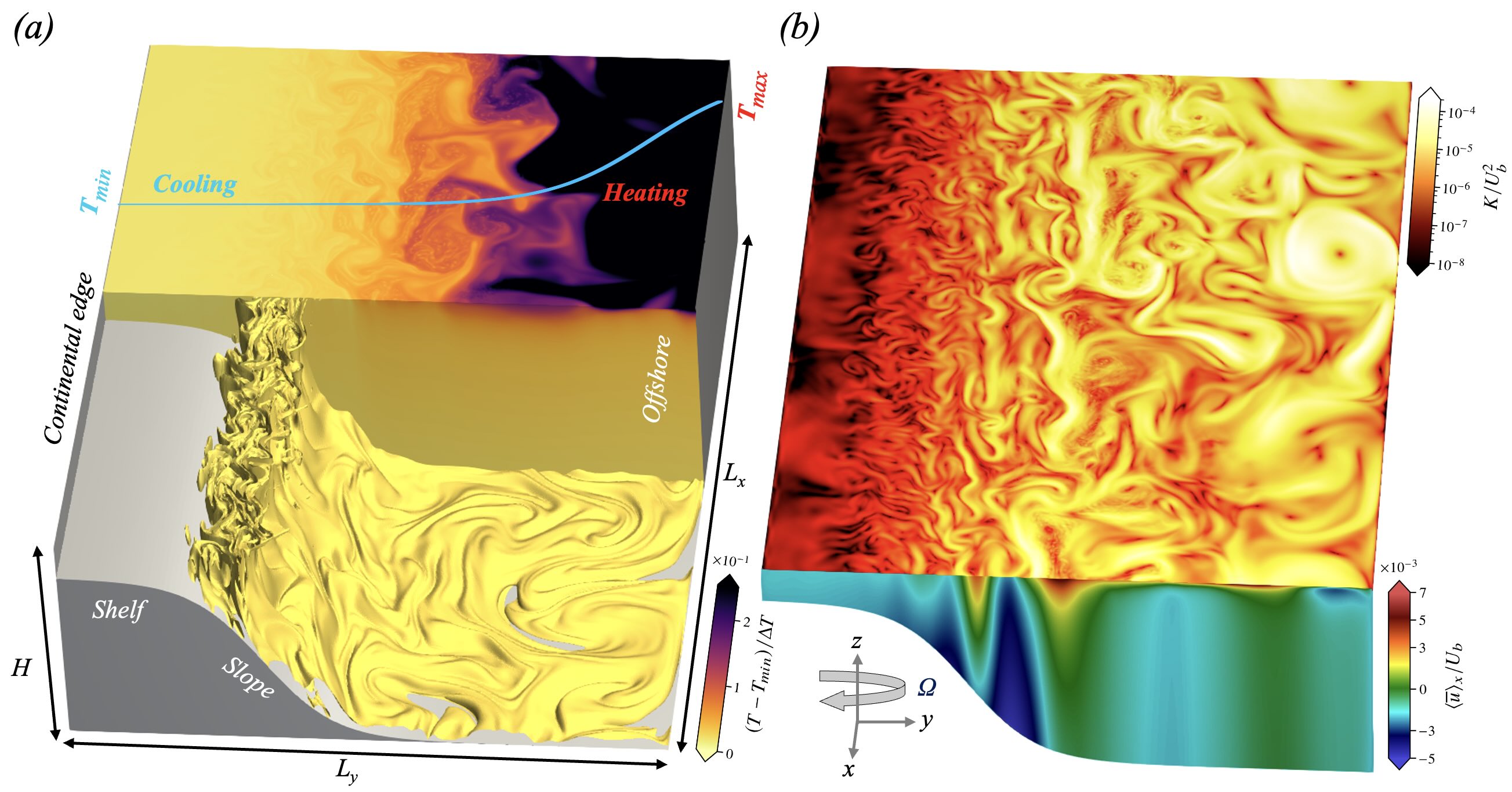}

\caption{
The idealised continental shelf--slope configuration and the key flow features considered in the present study for the intermediate-rotation case ($Ro=3.6\times10^{-2}$). $\mathit{(a)}$ Convection and dense-water formation driven by the imposed surface-temperature distribution. Three instantaneous isosurfaces of the nondimensional temperature, $\hat{T}=(T-T_{\min})/\Delta T$, at $\hat{T}=0.005$, $0.010$, and $0.019$ illustrate the three-dimensional structure of the convective flow and dense-water transport over the slope. Supplementary movie 1 shows the time evolution of the temperature isosurface at $\hat{T}=0.005$ for the
non-rotating and rotating cases. $\mathit{(b)}$ Instantaneous kinetic energy per unit mass in the upper part of the domain, normalised by the square of the buoyancy velocity scale, $K/U_b^2$, where $K=\frac{1}{2}\boldsymbol{u}\cdot\boldsymbol{u}$ and $U_b=(g\alpha\Delta T L)^{1/2}$ is the buoyancy velocity scale. The front face shows the time- and along-slope averaged along-slope velocity, $\langle \overline{u}\rangle_x$, highlighting the oppositely directed surface-intensified and slope-attached currents. Supplementary movie 2 shows the time evolution of the normalised kinetic energy, $K/U_b^2$, and along-slope averaged
vertical velocity, $\langle w\rangle_x/U_b$.
}
\label{fig:model}

\label{fig:model}
\end{figure}

Periodic boundary conditions are imposed in the along-slope direction $\hat{x}$. No-slip boundary conditions are applied at the sloping bottom, $\hat{z}=\hat{z}_b(\hat{y})$, and at the cross-slope boundaries, $\hat{y}=0$ and $\hat{y}=1$, such that $\hat{\boldsymbol{u}}=(\hat{u},\hat{v},\hat{w})=\boldsymbol{0}$, where $\hat{u}$, $\hat{v}$ and $\hat{w}$ are the along-slope, cross-slope and vertical velocity components, respectively. The upper boundary, $\hat{z}=H/L$, is impermeable and free-slip, with $\hat{w}=0$, $\partial\hat{u}/\partial\hat{z}=0$ and $\partial\hat{v}/\partial\hat{z}=0$. The sloping bottom and cross-slope boundaries are thermally insulated, such that $\boldsymbol{n}\cdot\hat{\boldsymbol{\nabla}}\hat{T}=0$, where $\boldsymbol{n}$ denotes the outward unit normal to the corresponding boundary and $\hat{T}$ is the nondimensional temperature. The model considers a single active scalar, temperature. Horizontal convection is driven by prescribing a cross-slope-varying temperature distribution, $\hat{T}(\hat{y})$, at the upper boundary, as shown in figure~\ref{fig:model}$(a)$. The imposed thermal forcing maintains relatively low surface temperatures over the continental shelf and slope and higher temperatures in the offshore region. The imposed temperature difference is defined as $\Delta T = T_{\max}-T_{\min}$.

The fluid is assumed to be incompressible and is described using the Boussinesq approximation. The continuity equation, the non-hydrostatic Navier--Stokes equations in a rotating reference frame, and the temperature conservation equation are written in nondimensional form as

\begin{subequations}
\label{eq:governing_equations}
\begin{align}
\hat{\boldsymbol{\nabla}}
\cdot\hat{\boldsymbol{u}}
&=0,
\label{eq:continuity}
\\
Pr^{-1}
\left(
\frac{D\hat{\boldsymbol{u}}}{D\hat{t}}
+
\hat{\boldsymbol{\nabla}}\hat{p}
\right)
&=
Ra\,\hat{T}\boldsymbol{e}_z
-
E^{-1}\boldsymbol{e}_z
\times\hat{\boldsymbol{u}}
+
\hat{\nabla}^{2}\hat{\boldsymbol{u}},
\label{eq:momentum}
\\
\frac{D\hat{T}}{D\hat{t}}
&=
\hat{\nabla}^{2}\hat{T}.
\label{eq:temperature}
\end{align}
\end{subequations}
Here, bold symbols denote vector quantities, and hats denote nondimensional variables. The quantity $\hat{p}$ denotes the nondimensional pressure deviation from the hydrostatic reference state.

The governing equations are nondimensionalised using the characteristic length scale $L=L_y$, where $L_y$ is the cross-slope extent of the domain, together with the thermal-diffusion time scale $L^2/\kappa$, velocity scale $\kappa/L$, imposed temperature difference $\Delta T$, and pressure scale $\rho_0\kappa^2/L^2$. Accordingly, $\boldsymbol{x}=L\hat{\boldsymbol{x}}$, $t=(L^2/\kappa)\hat{t}$, $\boldsymbol{u}=(\kappa/L)\hat{\boldsymbol{u}}$, $\hat{T}=(T-T_{s,\min})/\Delta T$, and $p=(\rho_0\kappa^2/L^2)\hat{p}$. Here, $\rho_0$ is the reference density, $\nu$ is the kinematic viscosity, $\kappa$ is the thermal diffusivity, and $\alpha$ is the thermal-expansion coefficient. The resulting dimensionless control parameters are the Rayleigh number $Ra$, the Prandtl number $Pr$, the Ekman number $E$, the vertical aspect ratio $A$, and the Rossby number $Ro$ \citep{whitehead2000circulation} These are defined as
\begin{align}
Ra &=
\frac{g\alpha\Delta T L^3}{\nu\kappa},
&
Pr &=
\frac{\nu}{\kappa},
\label{eq:ra_pr}
\\
E &=
\frac{\nu}{fL^2},
&
A &=
\frac{H}{L},
\label{eq:e_a}
\\
Ro &=
\frac{g\alpha\Delta T H}{f^2L^2}.
\label{eq:rossby_number}
\end{align}
The characteristic geostrophic velocity is $U_g=g\alpha\Delta T H/(fL)$, obtained by balancing the horizontal pressure gradient associated with the imposed temperature difference against the Coriolis acceleration. The Rossby number therefore measures the strength of the thermally driven geostrophic motion relative to the rotational velocity scale $fL$. The characteristic buoyancy velocity is $U_b=(g\alpha\Delta T L)^{1/2}$, with corresponding buoyancy time scale $\tau_b=L/U_b$ \citep{Vreugdenhil_Gayen_Griffiths_2019}{}. The non-rotating case is denoted by $Ro=\infty$.

The Rayleigh number associated with oceanic horizontal convection is estimated to range from $Ra=1.2\times10^{25}$ to $1.9\times10^{26}$ for the Southern Ocean \citep{sohail2018convection}, which remains beyond the range currently accessible to direct numerical simulation. Nevertheless, dynamically relevant transitions in horizontal convection can be investigated at lower Rayleigh numbers, provided that the relevant dynamical regime is captured. In particular, \citet{gayen2014stability} employed direct numerical simulations and large-eddy simulations to investigate the transition from laminar to turbulent horizontal convection over the range $Ra\sim10^8$--$10^{15}$ at $Pr=5$.

Therefore, the present study considers a computationally accessible high-Rayleigh-number regime within the range investigated by \citet{gayen2014stability}. The dimensional fluid properties and thermal forcing are $\nu=1.0\times10^{-6}\,\mathit{m^2\,s^{-1}}$, $\kappa=2.0\times10^{-7}\,\mathit{m^2\,s^{-1}}$, $g\alpha=8.0\times10^{-3}\,\mathit{m\,s^{-2}\,K^{-1}}$ and $\Delta T=24\,\mathit{K}$. The cross-slope domain length is $L_y=1.5\,\mathit{m}$ and, as defined above, is used as the characteristic length scale, $L\equiv L_y$. These parameters give $Pr=5$ and $Ra=3.24\times10^{12}$, while the domain geometry gives a vertical aspect ratio $A=H/L=0.4$. One non-rotating case and three rotating cases are considered. The rotating simulations correspond to Coriolis parameters $f=0.6$, $1.2$ and $2.4\,\mathit{s^{-1}}$. The associated Ekman and Rossby numbers are summarised in table~\ref{tab:simulation_parameters}. The Rayleigh number, Prandtl number, domain geometry and thermal forcing are held fixed across all simulations.
\begin{table}
\begin{center}
\begin{tabular}{lcc}
\hline
Case & $E$ & $Ro$ \\[3pt]
\hline
Non-rotating
& $\infty$
& $\infty$ \\

Weak rotation
& $7.41\times10^{-7}$
& $1.4\times10^{-1}$ \\

Intermediate rotation
& $3.70\times10^{-7}$
& $3.6\times10^{-2}$ \\

Strong rotation
& $1.85\times10^{-7}$
& $8.9\times10^{-3}$ \\
\hline
\end{tabular}
\caption{
Ekman number ($E$) and Rossby number ($Ro$) for the non-rotating and rotating simulations. The Rayleigh number, Prandtl number, domain geometry, and thermal forcing are held fixed across all cases.
}
\label{tab:simulation_parameters}
\end{center}
\end{table}

The governing equations~\eqref{eq:continuity}--\eqref{eq:temperature} are solved using the spectral-element code \textsc{Nek5000} \citep{fischer2008nek5000}. The computational mesh contains approximately $6.5\times10^5$ hexahedral spectral elements. Within each element, the velocity and temperature fields are represented by polynomial expansions of degree $N=7$, corresponding to $N+1=8$ Gauss--Lobatto--Legendre points in each coordinate direction and therefore approximately $6.5\times10^5(8^3)\simeq3.3\times10^8$ element-local collocation points in total. The mesh is strongly refined near the upper thermal boundary layer and the sloping bottom, where the largest velocity and temperature gradients occur. Time integration is performed using a third-order backward-differentiation scheme, with the nonlinear advection terms treated explicitly and the viscous and thermal-diffusion terms treated implicitly through Helmholtz-type solves. The time step is adjusted to maintain a target Courant number of $0.5$.

The adequacy of the numerical resolution is assessed using the local Batchelor microscale, defined as $\eta_b = Pr^ {-1/2}(\nu^3/\epsilon)^{1/4}$, where $\epsilon$ is the local kinetic energy dissipation. The local spacing between adjacent Gauss--Lobatto--Legendre points is compared with $\eta_b$ throughout the computational domain to verify that the dynamically relevant velocity and temperature scales are resolved. Numerical accuracy is additionally assessed using the global energy budgets. The residuals of the corresponding energy balances, together with the Batchelor-scale resolution assessments, are presented in Appendix~\ref{app:Energy_budget}.

All simulations are initialised with a uniform nondimensional temperature, $\hat{T}_{\mathit{ini}}=(T_{\mathit{ini}}-T_{min})/\Delta T=0.5$. A small-amplitude random perturbation is imposed on the initial velocity field to activate the development of the flow. Each simulation is allowed to evolve until a thermally equilibrated state is reached, as indicated by the temporal evolution of the volume-averaged temperature, volume-integrated kinetic energy and surface-integrated heat flux. Following equilibration, each simulation is continued for at least $600\,\tau_b$ in order to calculate the time-averaged fields. An overbar denotes averaging over this interval. Angle brackets with a subscript denote spatial averaging over the indicated coordinate. For example, the along-slope average of a quantity $\phi$ is defined as $\langle\phi\rangle_x=L_x^{-1}\int_0^{L_x}\phi(x,y,z,t)\,\mathit{d}x$. Consequently, $\langle\overline{\phi}\rangle_x$ denotes a quantity that is first averaged in time over the statistically stationary interval and then averaged in the along-slope direction.

\section{Results}
\label{sec:Results}
\subsection{Mean circulation and along-slope transport}
\label{sec:Mean_circulation}
\begin{figure}[h]
\centering
\includegraphics[width=\textwidth]{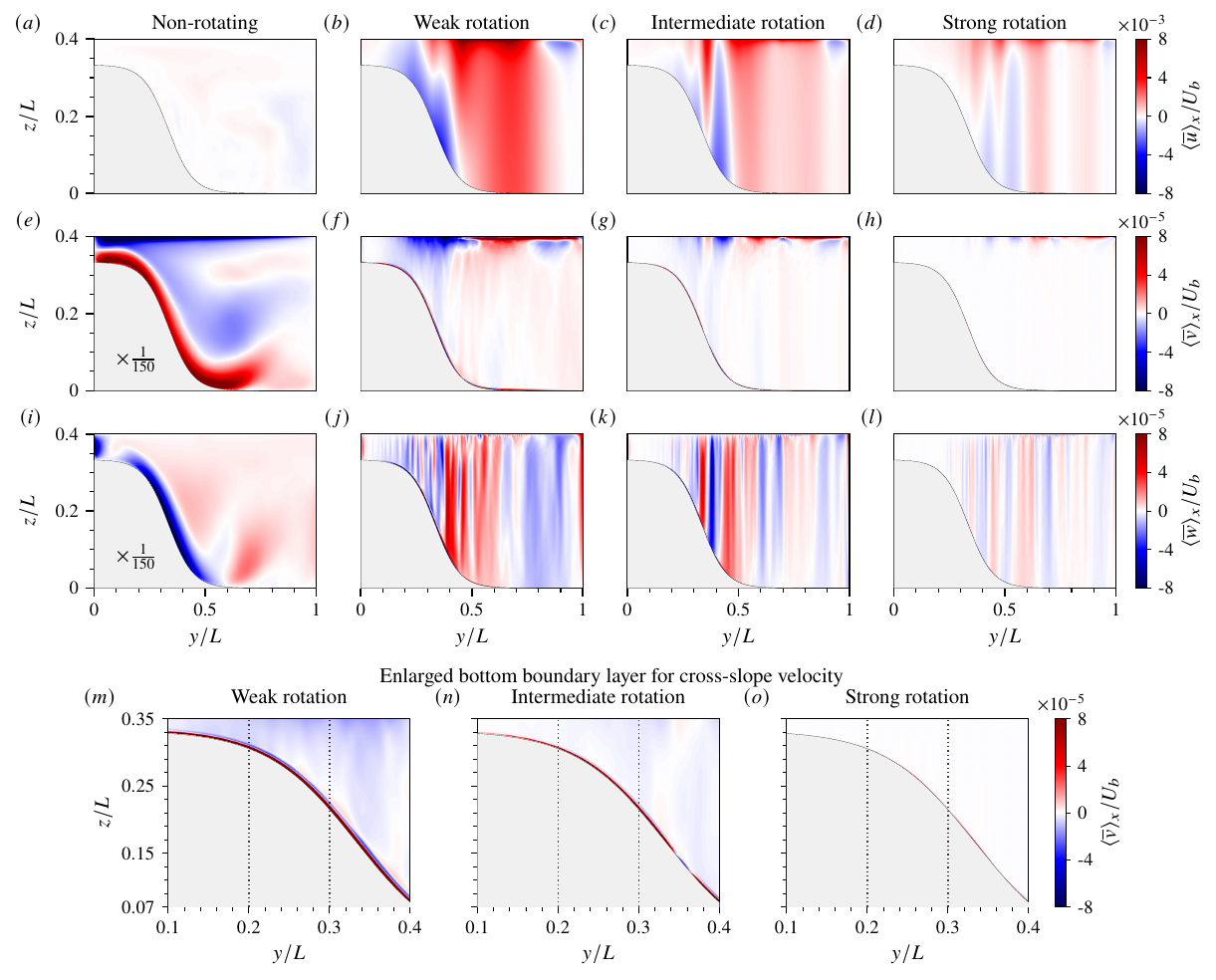}
\caption{
Time- and along-slope averaged velocity fields in the $(y,z)$ plane, evaluated over the statistically stationary interval. The columns correspond to the non-rotating case $\mathit{(a), (e), (i)}$, weak rotation with $Ro=1.4\times10^{-1}$ $\mathit{(b), (f), (j)}$, intermediate rotation with $Ro=3.6\times10^{-2}$ $\mathit{(c), (g), (k)}$, and strong rotation with $Ro=8.9\times10^{-3}$ $\mathit{(d), (h), (l)}$. The rows $\mathit{(a), (b), (c), (d)}$ show the along-slope velocity, $\langle\overline{u}\rangle_x$, $\mathit{(e), (f), (g), (h)}$ show the cross-slope velocity, $\langle\overline{v}\rangle_x$, and $\mathit{(i), (j), (k), (l)}$ show the vertical velocity, $\langle\overline{w}\rangle_x$. Panels $\mathit{(m), (n), (o)}$ provide enlarged views of the slope-attached bottom boundary layer in the cross-slope velocity field for the weak, intermediate, and strong rotation cases, respectively. The vertical dotted lines in panels $\mathit{(m), (n), (o)}$ indicate the cross-slope sampling locations at $y/L=0.2$ and $y/L=0.3$, used for the bottom Ekman-layer thickness and transport calculations. For visualisation, the cross-slope and vertical velocities in the non-rotating case are divided by $150$, as indicated in the corresponding panels. For the along-slope velocity, positive red values correspond to flow out of the page, and negative blue values to flow into the page.
}

\label{fig:mean_velocity_fields}
\end{figure}

We first examine how dense-water flow over the continental slope is modified by rotation and the subsequent dynamical response of the system, as illustrated in figure~\ref{fig:mean_velocity_fields}. The figure shows the transition from a non-rotating, gravity-driven circulation to a rotationally dominated flow, characterised by the along-slope, cross-slope and vertical velocity components, all averaged in time and along the slope direction, denoted by $\langle\overline{u}\rangle_x$, $\langle\overline{v}\rangle_x$ and $\langle\overline{w}\rangle_x$, respectively.
 
In the non-rotating case (figures~\ref{fig:mean_velocity_fields}$\mathit{(a,e,i)}$), the mean along-slope velocity is close to zero, with no coherent large-scale along-slope current (figure~\ref{fig:mean_velocity_fields}$\mathit{(a)}$). The cross-slope flow occupies a substantial fraction of the water column and extends far from the slope (figure~\ref{fig:mean_velocity_fields}$\mathit{(e)}$). Strong vertical motion accompanies this broad cross-slope flow, with downward motion concentrated along the slope and compensating upward motion farther offshore (figure~\ref{fig:mean_velocity_fields}$\mathit{(i)}$), allowing the gravity current to penetrate deeply and spread downslope.

Once rotation is introduced, in the weak-rotation case (figures~\ref{fig:mean_velocity_fields}$\mathit{(b,f,j)}$), two oppositely directed along-slope currents develop: a negative slope-attached current and a positive surface-intensified current extending into the offshore region (figure~\ref{fig:mean_velocity_fields}$\mathit{(b)}$). At this rotation rate, both currents penetrate through a substantial fraction of the water column. The cross-slope flow becomes localised near the topography (figure~\ref{fig:mean_velocity_fields}$\mathit{(f)}$), and a secondary near-surface flow close to the northern boundary is also present. The broad vertical-motion pattern of the non-rotating case is replaced by narrower regions of upward and downward velocity (figure~\ref{fig:mean_velocity_fields}$\mathit{(j)}$), including a bottom boundary-attached descending region and vertically coherent structures extending through much of the water column. Supplementary movie 3 shows the time evolution of the along-slope velocity field ($u$) over the slope region in the non-rotating and rotating cases.

At intermediate rotation (figures~\ref{fig:mean_velocity_fields}$\mathit{(c,g,k)}$), the positive along-slope current becomes increasingly confined to the upper part of the domain, while the negative current remains concentrated near the continental slope (figure~\ref{fig:mean_velocity_fields}$\mathit{(c)}$). The along-slope velocity field also becomes increasingly fragmented over the slope, forming alternating regions of positive and negative velocity. The cross-slope velocity becomes increasingly restricted to a thin slope-adjacent layer, while cross-slope motion in the interior is strongly suppressed (figure~\ref{fig:mean_velocity_fields}$\mathit{(g)}$). The secondary near-surface flow also decreases as the rotation increases. The vertical velocity structures become progressively narrower and more spatially confined (figure~\ref{fig:mean_velocity_fields}$\mathit{(k)}$).

At strong rotation (figures~\ref{fig:mean_velocity_fields}$\mathit{(d,h,l)}$), the positive along-slope current is further confined to the upper part of the domain, while the negative along-slope current becomes weaker and is concentrated slightly farther downslope, over the lower part of the continental slope (figure~\ref{fig:mean_velocity_fields}$\mathit{(d)}$). The along-slope velocity field remains fragmented over the slope, with alternating regions of positive and negative velocity, although the overall current intensity is reduced compared with the other rotating cases. The cross-slope velocity is restricted to a thin slope-adjacent layer, with cross-slope motion in the interior strongly suppressed (figure~\ref{fig:mean_velocity_fields}$\mathit{(h)}$). The vertical motions are also weaker and narrower than in the other rotating cases (figure~\ref{fig:mean_velocity_fields}$\mathit{(l)}$).

\begin{figure}[t]
\centering
\includegraphics[width=1.0\textwidth]{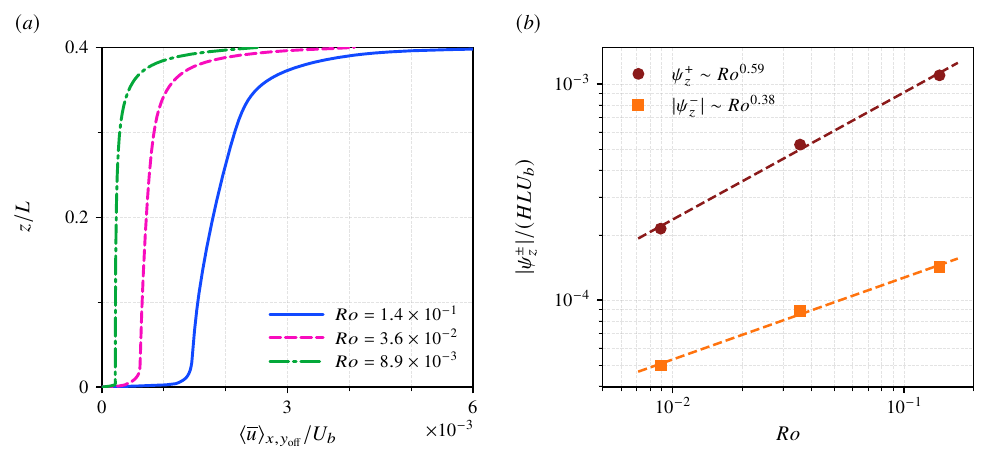}

\caption{
Effect of Rossby number on the vertical structure of the positive surface-intensified offshore along-slope current and the along-slope volume transports across the entire fluid cross-section. $\mathit{(a)}$ Vertical profiles of the time- and along-slope-averaged along-slope velocity, subsequently averaged in the cross-slope direction over the offshore region $0.6\leq y/L\leq0.9$ and normalised by the buoyancy velocity scale $U_b=(g\alpha\Delta T L)^{1/2}$. $\mathit{(b)}$ Positive along-slope transport, $\psi_z^{+}$ (circles), and magnitude of the negative along-slope transport, $|\psi_z^{-}|$ (squares), normalised by $HLU_b$, as functions of $Ro$. The transports are calculated over their respective positive and negative mean-velocity regions within the entire $(y,z)$ fluid cross-section. Here, $H$ is the maximum water-column depth and $L$ is the cross-slope domain length. Dashed lines indicate empirical power-law fits to the three rotating cases.
}
\label{fig:zonal_transport}
\end{figure}

To further examine the vertical structure of the positive, surface-intensified along-slope current, figure~\ref{fig:zonal_transport}$\mathit{(a)}$ shows vertical profiles of the time- and along-slope averaged along-slope velocity, subsequently averaged in the cross-slope direction over the offshore region $0.6\leq y/L\leq0.9$. The resulting profile is denoted by $\langle\overline{u}\rangle_{x,y_{\mathit{off}}}(z)$. These profiles provide a direct comparison of the magnitude and vertical distribution of the surface-intensified offshore current across the three rotating cases. The weakly rotating case ($Ro=1.4\times10^{-1}$) exhibits the strongest surface-intensified current, with its velocity profile extending over a greater fraction of the water column and a relatively broad vertical shear layer. As rotation increases ($Ro=3.6\times10^{-2}$ and $8.9\times10^{-3}$), the current progressively weakens, and the vertical shear becomes increasingly concentrated within a thinner layer near the upper boundary.

To quantify the volume transport associated with the oppositely directed along-slope currents, the time- and along-slope averaged velocity, $\langle\overline{u}\rangle_x(y,z)$, is integrated over the positive and negative velocity regions within the entire cross-slope--vertical $(y,z)$ fluid cross-section. The positive and negative along-slope transports are defined as
\begin{equation}
\psi_z^{\pm}
=
\iint_{\mathcal{A}^{\pm}}
\langle\overline{u}\rangle_x(y,z)\,
\mathit{d}y\,\mathit{d}z,
\qquad
\mathcal{A}^{\pm}
=
\left\{
(y,z)\in\mathcal{A}:
\langle\overline{u}\rangle_x(y,z)
\gtrless 0
\right\},
\label{eq:zonal_transport}
\end{equation}
where $\mathcal{A}$ denotes the entire fluid cross-sectional area in the $(y,z)$ plane. The subdomains $\mathcal{A}^{+}$ and $\mathcal{A}^{-}$ contain all locations where the time- and along-slope averaged velocity is positive and negative, respectively. The positive transport, $\psi_z^{+}$, is predominantly associated with the surface-intensified along-slope current, whereas the negative transport, $\psi_z^{-}$, is predominantly associated with the slope-attached current. Since $\psi_z^{-}<0$, its magnitude, $|\psi_z^{-}|$, is used when comparing the strengths of the two oppositely directed currents.

The variation of the along-slope transports with Rossby number is shown in figure~\ref{fig:zonal_transport}$\mathit{(b)}$. Both the positive transport, $\psi_z^{+}$, associated with the surface-intensified along-slope current, and the magnitude of the negative transport, $|\psi_z^{-}|$,
associated with the slope-attached current, decrease as $Ro$ decreases. The positive branch decreases more rapidly than the negative branch. Empirical power-law fits over the three rotating simulations give $\psi_z^{+}\propto Ro^{0.59}$ and $|\psi_z^{-}|\propto Ro^{0.38}$. Because these exponents are obtained from only three finite-$Ro$ simulations, they are interpreted as effective dependencies over the parameter range considered rather than as general scaling laws.

\subsection{Cross-slope transport and bottom boundary-layer contributions}
\label{sec:cross_slope_transport}

\begin{figure}[t]
\centering
\includegraphics[width=\textwidth]{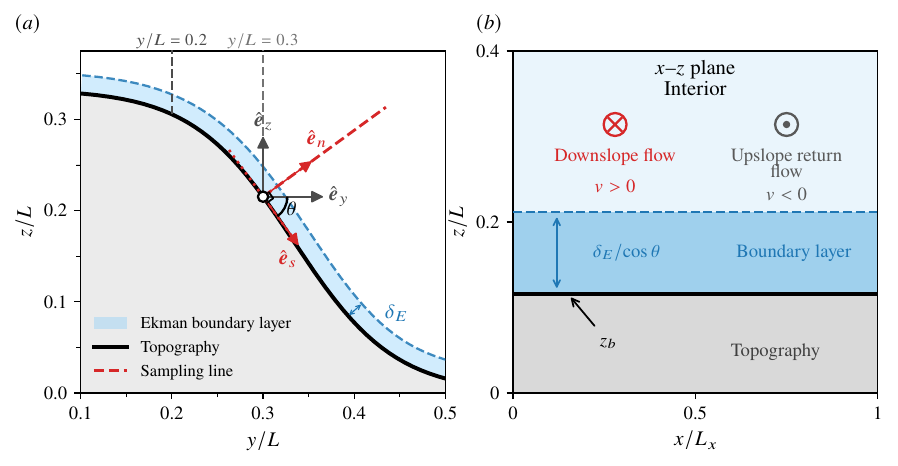}
\caption{Schematic of the slope-following coordinate system and the separation between bottom boundary-layer and interior transport. $\mathit{(a)}$ Local slope-parallel ($\hat{\boldsymbol e}_s$) and bottom-normal ($\hat{\boldsymbol e}_n$) directions, defined relative to the Cartesian $(\hat{\boldsymbol e}_y,\hat{\boldsymbol e}_z)$ basis. The angle $\theta$ is the local bottom inclination, the red dashed line indicates the bottom-normal direction and the blue shading denotes a bottom boundary layer of normal thickness $\delta_E$. Vertical dashed lines mark the sections at $y/L=0.2$ and $0.3$. $\mathit{(b)}$ Vertical $x$--$z$ section, showing the bottom boundary layer extending from the local bottom elevation $z_b(y)$ to $z_b(y)+\delta_E/\cos\theta$. Crosses and dots denote positive (downslope) and negative (upslope) cross-slope velocities, respectively. Both panels are schematic and not to scale.}
\label{fig:ekman_layer_geometry}
\end{figure}

The mean circulation described in \S~\ref{sec:Mean_circulation} includes a slope-attached current with a cross-slope component that can exchange fluid between the shelf and deeper regions. To identify the slope-attached bottom boundary layer, the velocity field is expressed in a local slope-following coordinate system, as illustrated in figure~\ref{fig:ekman_layer_geometry}. At the two prescribed positions, $y/L=0.2$ and $0.3$, the local bottom gradient is defined as $s_b=\mathit{d}z_b/\mathit{d}y$, with the inclination angle $\theta=\tan^{-1}(-s_b)$. The slope-parallel velocity is obtained by projecting the cross-slope and vertical velocity components onto the local downslope direction, $v^*=v\cos\theta-w\sin\theta$. The resulting velocity is averaged in time and subsequently along the periodic $x$-direction to obtain the mean slope-parallel velocity, $\langle\overline{v^*}\rangle_x$, as a function of the bottom-normal distance $n^{*}$. Details of the coordinate transformation are provided in Appendix~\ref{app:slope_coordinates}. 

The bottom boundary-layer thickness is diagnosed from the mean slope-parallel velocity profile, $\langle\overline{v^*}\rangle_x$, by first identifying the dominant near-bottom maximum in its absolute magnitude. Moving away from this maximum along the bottom-normal direction, the boundary-layer thickness, $\delta_E$, is determined by the first persistent zero crossing of the mean velocity. The resulting boundary-layer thicknesses are compared with the classical viscous Ekman scale, $\delta_{\mathit{E}}=(2\nu/f)^{1/2}$.

\begin{figure}[t]
\centering
\includegraphics[width=\textwidth]{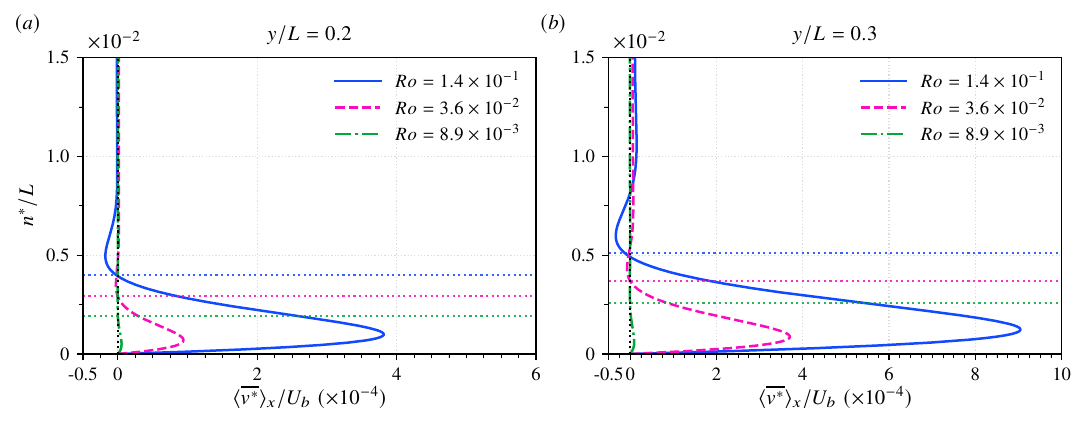}
\caption{Normalised time- and along-slope-averaged slope-parallel velocity, $\langle\overline{v^*}\rangle_x/U_b$, versus normalised bottom-normal distance, $n^{*}/L$, at $\mathit{(a)}$ $y/L=0.2$ and $\mathit{(b)}$ $y/L=0.3$. Blue solid, magenta dashed and green dash-dotted curves correspond to weak ($Ro=1.4\times10^{-1}$), intermediate ($Ro=3.6\times10^{-2}$) and strong ($Ro=8.9\times10^{-3}$) rotation, respectively. Coloured horizontal dotted lines mark the diagnosed boundary-layer thickness $\delta_E$ in each case.}
\label{fig:ekman_layer_velocity}
\end{figure}

\begin{figure}[t]
\centering
\includegraphics[width=\textwidth]{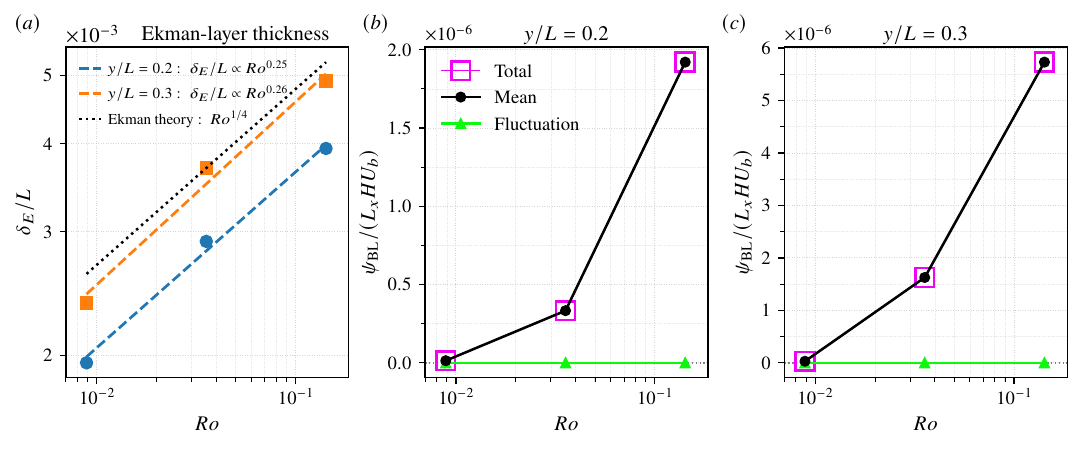}
\caption{
Bottom Ekman boundary-layer thickness and associated cross-slope volume transport as functions of Rossby number. $\mathit{(a)}$ Normalised Ekman-layer thickness, $\delta_E/L$, as a function of $Ro$ at $y/L=0.2$ (blue circles) and $y/L=0.3$ (orange squares). Dashed lines represent the power-law fits, $\delta_E/L\propto Ro^{0.25}$ and $\delta_E/L\propto Ro^{0.26}$, respectively. The black dotted line indicates the theoretical $Ro^{1/4}$ dependence. $\mathit{(b,c)}$ Total (magenta squares), mean (black circles) and fluctuation (green triangles) contributions to the signed cross-slope volume transport of fluid colder than $\widehat{T}_c=0.006$ within the diagnosed bottom boundary layer at $y/L=0.2$ and $y/L=0.3$, respectively. The mean and total transport curves nearly coincide because the fluctuation contribution is comparatively small.
}
\label{fig:ekman_thickness_transport}
\end{figure}

The slope-parallel velocity profiles shown in figure~\ref{fig:ekman_layer_velocity} indicate that the slope-parallel current is concentrated near the bottom and becomes confined to a thinner layer with increasing rotation. The diagnosed thickness is larger at $y/L=0.3$ than at $y/L=0.2$. Figure~\ref{fig:ekman_thickness_transport}$\mathit{(a)}$ quantifies this trend, with fitted dependences of $\delta_E/L\propto Ro^{0.25}$ and $\delta_E/L\propto Ro^{0.26}$ at $y/L=0.2$ and $0.3$, respectively. These exponents are close to the classical viscous Ekman scaling, $\delta_{\mathit{E}}=(2\nu/f)^{1/2}$, which, using the definition of the Rossby number in equation~\eqref{eq:rossby_number}, gives $\delta_{\mathit{E}}/L\propto Ro^{1/4}$.

However, the mean velocity field and boundary-layer thickness alone do not determine the transport of water in different temperature classes, because velocity fluctuations can be correlated with the evolving temperature distribution. Motivated by the temperature-coordinate framework of \citet{ferrari2011processes}, we quantify the cumulative cross-slope volume transport of colder fluid and separate the contributions associated with the time-mean velocity and its fluctuations. The diagnosed boundary-layer thickness is subsequently used to determine the fraction of gross downslope cold-water transport carried within it.

The transport is evaluated through vertical $x$--$z$ sections at $y/L=0.2$ and $0.3$, as indicated by the vertical dashed lines in figure~\ref{fig:ekman_layer_geometry}$\mathit{(a)}$. Each section extends from the local bottom elevation $z_b(y)$ to the upper boundary at $z/L=0.4$. At each section, the instantaneous cross-slope velocity $v(x,y,z,t)$ and temperature $T(x,y,z,t)$ are used to evaluate the transport before temporal averaging. The time-averaged cumulative cross-slope volume transport of fluid colder than a prescribed normalised temperature threshold $\widehat{T}_c$ is defined as
\begin{equation}
\psi_{\mathit{cr}}(y,\widehat{T}_c)
=
\frac{1}{\Delta t}
\iiint_{\widehat{T}\leq\widehat{T}_c}
v(x,y,z,t)\,\mathit{d}x\,\mathit{d}z\,\mathit{d}t,
\label{eq:cold_water_transport}
\end{equation}
where $\widehat{T}_c=(T_c-T_{\min})/\Delta T$. The temperature condition is applied to the instantaneous field; thus the portion of the section included in the integral can vary with time. The time-averaged transport is evaluated over an averaging interval $\Delta t=600\tau_b$ and has units of volume transport.

To evaluate the cumulative transport through each vertical $x$--$z$ section, the temperature range is divided into $8000$ classes, with the number of classes selected based on a sensitivity analysis. The instantaneous cross-slope volume transport associated with each class is accumulated from the coldest to the warmest fluid. When all temperature classes are included, the cumulative transport recovers the full-column net cross-slope volume flux, which vanishes in the present closed configuration.

The total transport can be decomposed into contributions from the time-mean flow and temporal fluctuations. Writing $v=\overline{v}+v'$, where the overbar denotes temporal averaging over the same interval $\Delta t$, gives
\begin{equation}
\psi_{\mathit{cr}}=\psi_{\mathit{cr}}^{\mathit{m}}+\psi_{\mathit{cr}}^{\mathit{e}},
\label{eq:cross_slope_decomposition}
\end{equation}
where
\begin{subequations}
\begin{align}
\psi_{\mathit{cr}}^{\mathit{m}}
&=
\frac{1}{\Delta t}
\iiint_{\widehat{T}\leq\widehat{T}_c}
\overline{v}(x,y,z)\,\mathit{d}x\,\mathit{d}z\,\mathit{d}t,
\label{eq:cross_slope_mean}
\\
\psi_{\mathit{cr}}^{\mathit{e}}
&=
\frac{1}{\Delta t}
\iiint_{\widehat{T}\leq\widehat{T}_c}
v'(x,y,z,t)\,\mathit{d}x\,\mathit{d}z\,\mathit{d}t.
\label{eq:cross_slope_mean_eddy}
\end{align}
\end{subequations}
Here, $\psi_{\mathit{cr}}^{\mathit{m}}$ represents transport by the time-mean velocity through the time-varying cold-water region, whereas $\psi_{\mathit{cr}}^{\mathit{e}}$ captures the contribution of velocity fluctuations correlated with that region.

\begin{figure}[t]
\centering
\includegraphics[width=\textwidth]{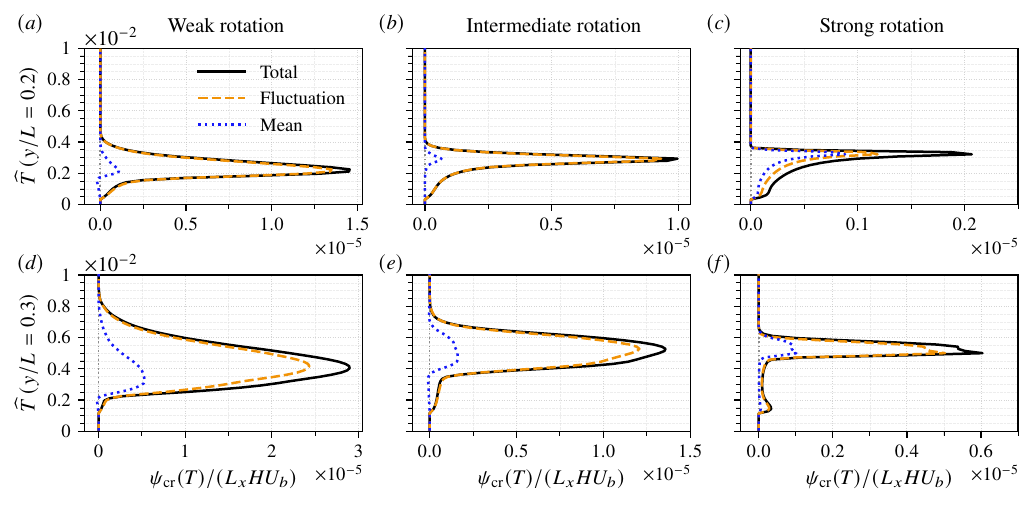}
\caption{Cumulative cross-slope volume transport in temperature coordinates, $\psi_{\mathit{cr}}/(L_xHU_b)$, through full vertical sections at $y/L=0.2$ (top row) and $0.3$ (bottom row). Columns correspond to $\mathit{(a),(d)}$ weak ($Ro=1.4\times10^{-1}$), $\mathit{(b),(e)}$ intermediate ($Ro=3.6\times10^{-2}$) and $\mathit{(c),(f)}$ strong ($Ro=8.9\times10^{-3}$) rotation. Black solid, blue dotted and orange dashed curves denote total, mean and fluctuation contributions, respectively. Positive transport is directed downslope. At each temperature threshold $\widehat{T}_c$, the cumulative transport includes all fluid satisfying $\widehat{T}\leq\widehat{T}_c$.}
\label{fig:cumulative_cross_slope_transport}
\end{figure}

Figure~\ref{fig:cumulative_cross_slope_transport} shows the cumulative cross-slope volume transport for the three rotating cases at $y/L=0.2$ and $0.3$. At both locations, the cumulative transport initially increases with the temperature, reaches a positive maximum and subsequently decreases towards zero as warmer temperature classes are included. The initial increase indicates a net downslope transport associated with colder fluid, while the subsequent decrease indicates a compensating net upslope contribution from relatively warmer fluid. The maximum cumulative transport decreases progressively with increasing rotation at both locations. For a given rotation rate, the maximum cumulative transport is greater at $y/L=0.3$ than at $y/L=0.2$ and occurs at a higher normalised temperature threshold.

The mean--fluctuation decomposition reveals that the fluctuation contribution accounts for most of the cumulative transport in all three rotating cases. At $y/L=0.2$, the total and fluctuation contributions nearly coincide under weak and intermediate rotation, while the mean contribution becomes more important under strong rotation. At $y/L=0.3$, the mean contribution remains smaller than the fluctuation contribution across all three cases, although its magnitude varies with rotation. 

At the two cross-slope locations, $y/L=0.2$ and $0.3$, the vertical thickness corresponding to the diagnosed bottom-normal boundary-layer thickness $\delta_E$ is given by $\delta_E/\cos\theta$, as illustrated in figure~\ref{fig:ekman_layer_geometry}$\mathit{(b)}$. The bottom boundary-layer contribution to the cross-slope transport is then evaluated using the same formulation as for the full water column, but with the vertical integration restricted to the corresponding boundary-layer extent at each location. The total, mean and fluctuation contributions are obtained by applying this restriction to equations~(\ref{eq:cold_water_transport})--(\ref{eq:cross_slope_mean_eddy}). To quantify the fraction of downslope dense-water transport carried within the bottom boundary layer, only the positive cross-slope velocity contributions are considered. The corresponding fraction is defined as $\mathcal F_{\mathit{BL}}=\psi_{\mathit{BL}}/\psi_{\mathit{cr}}$, where $\psi_{\mathit{BL}}$ and $\psi_{\mathit{cr}}$ represent the downslope transport of fluid colder than the prescribed temperature threshold within the bottom boundary layer and the full water column, respectively. A sensitivity analysis was performed over the normalised temperature threshold range $0.002\leq\widehat{T}_c\leq0.012$. A common threshold of $\widehat{T}_c=0.006$ was selected for all rotating cases at both cross-slope locations, $y/L=0.2$ and $0.3$, beyond which changes in the diagnosed dense-water transport were negligible.

Figure~\ref{fig:ekman_thickness_transport}$\mathit{(b,c)}$ shows the cross-slope volume transport within the bottom boundary layer and its mean and fluctuation contributions. The magnitude of the total transport decreases with increasing rotation at both locations. At both $y/L=0.2$ and $0.3$, the mean contribution nearly coincides with the total transport, while the fluctuation contribution remains negligible across all three rotating cases.

The fraction of the total downslope dense-water transport carried within the bottom boundary layer is evaluated for the prescribed temperature threshold $\widehat{T}_c=0.006$. At $y/L=0.2$, the contribution decreases from $3.9,\%$ under weak rotation to $2.5,\%$ and $1.1,\%$ under intermediate and strong rotation, respectively. At $y/L=0.3$, the corresponding contributions are $4.1,\%$, $2.8,\%$ and $0.8,\%$. Overall, the bottom boundary-layer contribution remains below $5,\%$ of the total downslope dense-water transport across all three rotating cases at both sampling locations. 

These results indicate that the bottom boundary layer makes only a small contribution to the net downslope transport of colder water, despite the dominance of the mean flow within this layer.

\subsection{Residual overturning circulation }
\label{sec:Residual_overturning_circulation}

The along-slope and cross-slope transport diagnostics presented in sections~\ref{sec:cross_slope_transport} and~\ref{sec:Mean_circulation} characterise the circulation and associated volume transport, but do not resolve the vertical overturning across different density classes. To examine the density-dependent overturning circulation, we diagnose the residual streamfunction in depth--density coordinates. This formulation quantifies the cumulative vertical volume transport associated with different density classes at each depth, incorporating contributions from both the mean flow and time-dependent fluctuations. Following \citet{nurser2004isopycnal}, the residual streamfunction is defined as
\begin{equation}
\psi_{r}(\rho^{*},z)
=
-\frac{1}{\Delta t}
\iiint_{\rho \ge \rho^{*}}
w(x,y,z,t)\,
dx\,dy\,dt,
\label{eq:depth_density_streamfunction}
\end{equation}
where $w$ is the vertical velocity, $\rho(x,y,z,t)=\rho_0[1-\alpha(T(x,y,z,t)-T_{\min})]$ is the instantaneous density determined from the linear equation of state, $\rho_0=1$ is the reference density, $\alpha$ is the thermal expansion coefficient, $T_{\min}$ is the reference minimum temperature, $\rho^{*}$ is the prescribed density, and $\Delta t=600\tau_b$ is the averaging period. At each depth, the integration is performed over the horizontal section and times for which $\rho\geq\rho^{*}$. The residual streamfunction therefore represents the cumulative vertical volume transport of fluid denser than $(\rho^{*})$. The residual streamfunction therefore represents the cumulative vertical volume transport of fluid denser than $\rho^{*}$.

For visualisation, the depth--density streamfunction is mapped onto an equivalent cross-slope coordinate following \citet{nurser2004isopycnal}. The cumulative horizontal area occupied by fluid denser than $\rho^{*}$ at each depth is mapped into an equivalent cross-slope width,
\begin{equation}
y(\rho^{*},z)
=
\frac{1}{L_{x}\,\Delta t}
\iiint_{\rho \ge \rho^{*}}
dx\,dy\,dt,
\end{equation}
The integration is restricted to the fluid region at each depth, accounting for the sloping topography. The resulting equivalent cross-slope coordinate preserves the cumulative horizontal area occupied by fluid denser than the chosen density threshold. The residual overturning circulation is then presented in equivalent cross-slope--depth $(y,z)$ coordinates.

\begin{figure}[h]
\centerline{\includegraphics[width=1.0\textwidth]{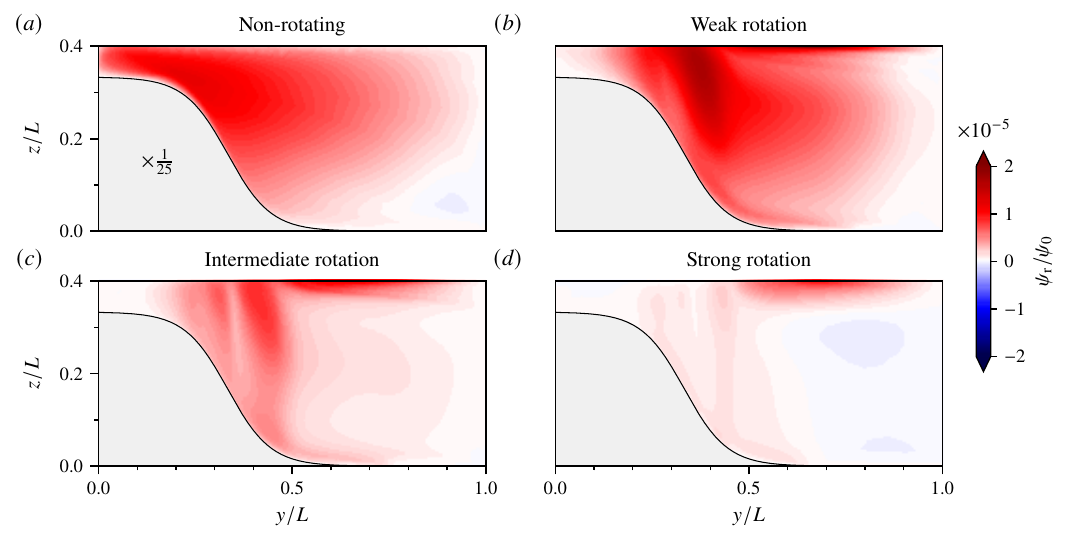}}
\caption{
Residual mean overturning circulation in depth--latitude space for $\mathit{(a)}$ Non-rotating case, $\mathit{(b)}$ $Ro=1.4\times10^{-1}$, $\mathit{(c)}$ $Ro=3.6\times10^{-2}$, and $\mathit{(d)}$ $Ro=8.9\times10^{-3}$. The colour bar shows the normalized overturning transport, with the streamfunction scaled by the characteristic transport $ \psi_{0}=L_{x} L (g\alpha\Delta T L)^{1/2}$. The non-rotating case is scaled by a factor of $1/25$ for visualization. Positive (red) and negative (blue) values correspond to anticlockwise and clockwise overturning circulation, respectively.
}
\label{fig:Residual_streamfunction}
\end{figure}

\begin{figure}[h]
\centering
\includegraphics[width=1.0\textwidth]{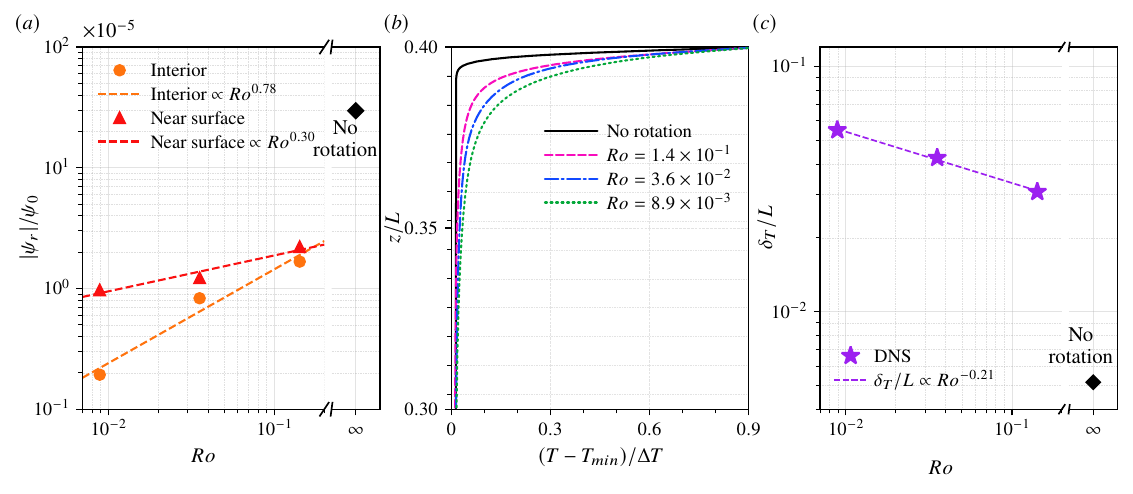}
\caption{
Effect of rotation on the residual overturning circulation and upper thermal boundary layer. $\mathit{(a)}$ Maximum magnitude of the normalised residual streamfunction, $|\psi_r|/\psi_0$, for the interior overturning cell (orange circles) and near-surface overturning cell (red triangles). Dashed lines indicate the corresponding empirical power-law fits. $\mathit{(b)}$ Vertical profiles of the normalised temperature, $(T-T_{s,\min})/\Delta T$, near the upper boundary for the non-rotating (black solid line), weakly rotating (magenta dashed line), intermediately rotating (blue dash-dotted line) and strongly rotating (green dotted line) cases. $\mathit{(c)}$ Normalised thermal boundary-layer thickness, $\delta_T/L$, as a function of Rossby number, $Ro$. Purple stars represent the rotating DNS results, and the purple dashed line indicates the corresponding empirical power-law fit. The non-rotating values ($Ro=\infty$) are indicated separately by black diamonds in panels $\mathit(a)$ and $\mathit(c)$ and are excluded from the power-law fits.
}
\label{fig:masstransport}
\end{figure}

The residual circulation transitions from domain-scale overturning to weaker, slope-trapped transport as the Rossby number decreases, as shown in figure~\ref{fig:Residual_streamfunction}. In the non-rotating case (figure~\ref{fig:Residual_streamfunction}$\mathit{(a)}$), a single positive cell occupies nearly the full water-column depth and extends across much of the domain, connecting the surface buoyancy forcing region to the bottom boundary. Under weak rotation (figure~\ref{fig:Residual_streamfunction}$\mathit{(b)}$), the overturning cell contracts horizontally and becomes concentrated above the continental slope. The dominant cell remains vertically extended, but its transport is organised within a narrow region over the slope rather than as a domain-wide circulation. At intermediate rotation (figure~\ref{fig:Residual_streamfunction}$\mathit{(c)}$), the domain interior transport is strongly reduced, and the dominant overturning becomes confined to the continental slope. Dense water formed over the shelf is transported primarily within a slope-adjacent circulation, with substantially weaker exchange between the shelf, slope, and deeper interior. Under the strongest rotation (figure~\ref{fig:Residual_streamfunction}$\mathit{(d)}$), the residual transport weakens further and becomes restricted mainly to the upper continental slope. Only very weak positive and negative recirculating cells remain in the domain interior, and these do not form a coherent connection between the surface-forcing region and the deep interior. All rotating cases also contain a thin secondary overturning cell near the upper boundary. This cell is absent in the non-rotating case and weakens as $Ro$ decreases.

Following the qualitative description of the residual circulation in figure~\ref{fig:Residual_streamfunction}, its variation with Rossby number is quantified in figure~\ref{fig:masstransport}. The maximum residual streamfunction values are extracted from two distinct overturning regions: the interior circulation and the near-surface circulation. Because these circulations are driven by the surface buoyancy forcing, their strengths are also examined together with the thermal boundary-layer thickness. This comparison links the changes in residual overturning strength to the rotational modification of the near-surface thermal structure.

The magnitude of the residual overturning decreases with decreasing Rossby number for both the interior and near-surface circulations. Based on the three rotating simulations, the interior overturning follows $\psi_{\mathit{int}}/\psi_0\propto Ro^{0.78}$, whereas the near-surface overturning exhibits a weaker dependence, $\psi_{\mathit{surf}}/\psi_0\propto Ro^{0.30}$, as shown in figure~\ref{fig:masstransport}$\mathit{(a)}$. The non-rotating value is shown separately and is excluded from the power-law fits.
 
The horizontally averaged temperature profiles in figure~\ref{fig:masstransport}$\mathit{(b)}$ show that the upper thermal boundary layer penetrates progressively farther into the fluid as $Ro$ decreases. Across the three rotating cases, the diagnosed boundary-layer thickness increases with decreasing Rossby number and is well described by the empirical dependence $\delta_T/L\propto Ro^{-0.209}$, as shown in figure~\ref{fig:masstransport}$\mathit{(c)}$. The non-rotating value is shown separately and is excluded from the fit. Overall, these results demonstrate that increasing rotation weakens the large-scale circulation while promoting the thickening of the upper thermal boundary layer.  

\subsection{Topographic Rossby waves}
\label{sec:trw}

Topographic Rossby waves (TRWs) are low-frequency motions supported by a potential-vorticity gradient associated with variations in water-column depth over sloping topography \citep{pedlosky1967laboratory}. In the present configuration, the continental slope provides a strong topographic potential-vorticity gradient, while surface buoyancy forcing generates the background stratification and along-slope current. We therefore examine whether the low-frequency oscillations observed in the DNS results exhibit the dispersion characteristics and spatial coherence expected of topographically controlled Rossby waves. For a continuously stratified fluid over a uniform, gently sloping bottom, the linear TRW dispersion relation may be written in the laboratory frame as \citep{pedlosky1987}
\begin{equation}
\omega_{lab}
=
\overline{U}k
+
\frac{N s k}{K}
\coth
\left(
\frac{N D K}{f}
\right),
\qquad
K
=
\left(
k^2+\ell^2
\right)^{1/2},
\label{eq:trw_dispersion}
\end{equation}
where $k$ and $\ell$ are the dimensional along-slope and cross-slope wavenumbers, respectively, $N$ is the buoyancy frequency, and $\overline{U}$ is the representative background along-slope velocity. The local water-column depth is given by $D(y)=H-z_b(y)$, where $H$ is the elevation of the top surface. The first term on the right-hand side of equation~\eqref{eq:trw_dispersion} represents the Doppler shift associated with the background along-slope current, while the second term represents the intrinsic TRW frequency.

The linear TRW dispersion relation is derived for an idealised, gently sloping bottom with uniform background stratification, whereas the present configuration exhibits spatially varying topography and stratification. Following previous applications of locally evaluated TRW theory over nonuniform bathymetry \citep[e.g.][]{guo2024seamount_trw}, we divide the continental slope into an upper-slope region ($0.20\leq y/L\leq0.30$) and a lower-slope region ($0.30<y/L\leq0.45$). The corresponding mean bottom gradients are $0.90$ and $1.19$, respectively, representing substantial departures from the classical gentle-slope approximation. The linear dispersion relation is therefore used as a local theoretical reference to assess whether the observed low-frequency variability exhibits characteristics of TRW dynamics, rather than as an exact quantitative prediction for the simulated flow.

The background stratification is quantified using the squared buoyancy frequency, $N^2(y,z)=g\alpha\,\partial\langle\overline{T}\rangle_x/\partial z$, where $\langle\overline{T}\rangle_x$ is the time- and along-slope-averaged temperature field. Positive values of $N^2$ indicate stable stratification, whereas $N^2<0$ indicates unstable stratification and identifies regions susceptible to convection under the imposed surface buoyancy forcing. For the theoretical comparison, a representative buoyancy frequency is selected from the stably stratified interior of each slope region, excluding locations where $N^2\leq0$. The adopted values are reported in table~\ref{tab:trw_parameters}.
\begin{table}
\centering
\begin{tabular}{lcccc}
\hline
$Ro$ & $N/N_b$ & $D/L$ & $|s|$ & $|\overline{U}|/U_b$ \\
\hline
\multicolumn{5}{l}{\textit{Upper slope}} \\
$1.42\times10^{-1}$ & 0.08 & 0.13 & 0.90 & $1.82\times10^{-3}$ \\
$3.56\times10^{-2}$ & 0.07 & 0.13 & 0.90 & $5.69\times10^{-4}$ \\
$8.89\times10^{-3}$ & 0.02 & 0.13 & 0.90 & $1.45\times10^{-5}$ \\
\hline
\multicolumn{5}{l}{\textit{Lower slope}} \\
$1.42\times10^{-1}$ & 0.06 & 0.3 & 1.19 & $1.81\times10^{-3}$ \\
$3.56\times10^{-2}$ & 0.06 & 0.3 & 1.19 & $1.32\times10^{-3}$ \\
$8.89\times10^{-3}$ & 0.04 & 0.3 & 1.19 & $3.62\times10^{-4}$ \\
\hline
\end{tabular}
\caption{ Representative nondimensional background parameters used to evaluate the linear topographic Rossby-wave dispersion relation over the upper- and lower-slope regions. Here $N/N_b$ is the normalised buoyancy frequency, $D/L$ is the normalised mean water-column depth, $|s|$ is the magnitude of the mean bottom gradient, and $|\overline{U}|/U_b$ is the magnitude of the representative background along-slope velocity, normalised by the buoyancy velocity scale. The background velocities are obtained from the time- and along-slope-averaged velocity field within the corresponding slope regions.
}
\label{tab:trw_parameters}
\end{table}

\begin{figure}
\centering
\includegraphics[width=\textwidth]{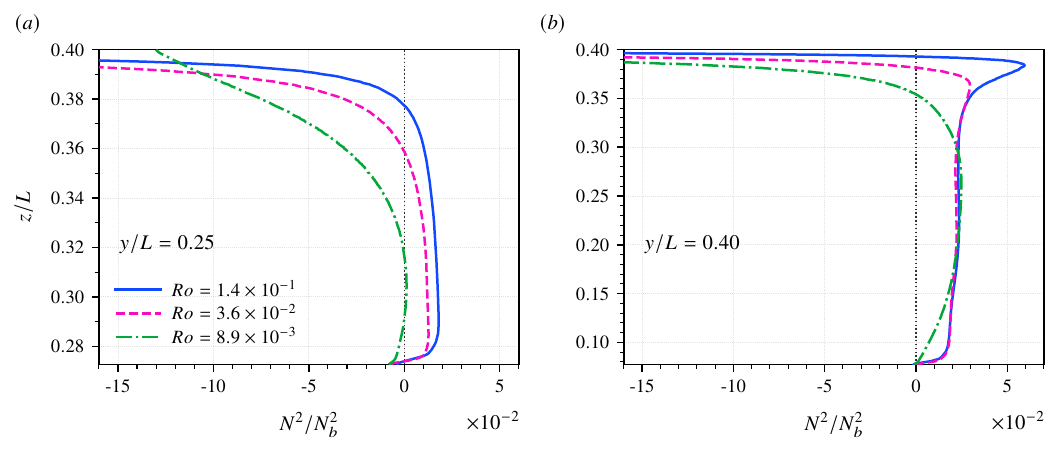}
\caption{
Vertical profiles of the squared buoyancy frequency, \(N^2\), normalised by \(N_b^2=g\alpha\Delta T/L\), at the corresponding locations within the upper- and lower-slope regions used for the TRW analysis: \(\mathit{(a)}\) \(y/L=0.25\), within \(0.20\leq y/L\leq0.30\), and \(\mathit{(b)}\) \(y/L=0.40\), within \(0.30<y/L\leq0.45\). Profiles are shown for the three rotating cases and are calculated from the time- and along-slope-averaged temperature field, \(\langle\overline{T}\rangle_x\). At each cross-slope location, the profile extends from the local sloping bottom, \(z=z_b(y)\), to the upper boundary, \(z=H\). The vertical dotted lines indicate neutral stratification, \(N^2=0\). The different vertical ranges of the two panels reflect the variation in local water-column depth over the continental slope.
}
\label{fig:N2_profiles}
\end{figure}

To characterise the background stratification in these two regions, figure~\ref{fig:N2_profiles} shows vertical profiles of $N^2/N_b^2$ at representative locations, $y/L=0.25$ and $0.40$, for the upper and lower slopes, respectively. Over the upper slope, represented by $y/L=0.25$ in figure~\ref{fig:N2_profiles}$\mathit{(a)}$, the interior is weakly stratified, while $N^2$ becomes negative immediately beneath the upper boundary, indicating a convectively unstable surface layer. As $Ro$ decreases, this unstable layer extends progressively further into the fluid, while the underlying stable stratification weakens. In the strongest-rotation case, $Ro=8.9\times10^{-3}$, the stable stratification in the lower interior is substantially weakened, with $N^2$ approaching zero over much of this region. Over the lower slope, represented by $y/L=0.40$ in figure~\ref{fig:N2_profiles}$\mathit{(b)}$, all three rotating cases exhibit a thin convectively unstable surface layer overlying a subsurface maximum in $N^2$. As $Ro$ decreases, the subsurface positive stratification weakens and is displaced deeper into the water column, while the negative-$N^2$ surface layer extends further into the fluid.

To examine the unsteady motions over the continental slope, the along-slope velocity is sampled on a topography-following surface,
\begin{equation}
u_s(x,y,t)
=
u\bigl(x,y,z_s(y),t\bigr),
\qquad
z_s(y)=z_b(y)+\delta_z(y),
\label{eq:trw_sample_surface}
\end{equation}
where $\delta_z(y)$ is the vertical distance above the local bottom topography. The sampling surface is selected to remain above the outer edge of the bottom Ekman layer, thereby reducing the direct contribution of the viscous near-bottom flow to the spectral analysis. The same sampling criterion is applied consistently to the three rotating cases.

The spectral analysis is performed separately over the upper- and lower-slope regions defined above. At each cross-slope sampling location, $y=y_j$, the along-slope velocity is collected over the full periodic domain and the analysis interval. After removing the temporal and along-slope means, a two-dimensional Fourier transform is performed in the along-slope and temporal directions. The spectral power is calculated separately at each cross-slope sampling location and subsequently averaged over the corresponding slope region. This procedure is similar to the spectral averaging approach of \citet{wang2010}. Averaging the squared Fourier amplitudes rather than the complex coefficients avoids cancellation arising from cross-slope variations in wave phase. The regionally averaged spectral power is given by
\begin{equation}
P(k^*,\omega)
=
\frac{1}{N_y}
\sum_{j=1}^{N_y}
\frac{
\left|\widehat{u'_j}(k^*,\omega)\right|^2
}{
U_b^2
},
\label{eq:regional_spectral_power}
\end{equation}
where $N_y$ is the number of cross-slope sampling locations within the respective upper- or lower-slope region, $\widehat{u'_j}$ is the two-dimensional Fourier coefficient of the along-slope velocity fluctuation at the $j$th sampling location, obtained by Fourier transformation in the along-slope direction and time, and $\omega$ is the intrinsic angular frequency after Doppler correction. The Fourier transform of the DNS velocity fluctuations initially provides frequencies in the laboratory frame. To compare the resulting spectra with the intrinsic TRW dispersion relation, the laboratory-frame frequency is Doppler-corrected using $\omega=\omega_{\mathit{lab}}-\overline{U}k$, where $\omega_{\mathit{lab}}$ is the laboratory-frame angular frequency. This correction shifts the frequency coordinate without changing the spectral power.

The resulting spectra are presented in terms of the dimensionless along-slope Fourier mode $k^*$ and the normalised intrinsic frequency $|\omega|/N_b$, allowing comparison with the intrinsic linear TRW dispersion relation. The temporal record consists of approximately $340$ snapshots spanning $600$ buoyancy time scales, and the lower limit of the plotted frequency range is set to $|\omega|/N_b=0.009$ to exclude frequencies close to zero while retaining the low-frequency variability.

The theoretical dispersion curves are evaluated separately for each Rossby number and slope region using the corresponding background parameters and an effective cross-slope wavenumber diagnosed from the DNS. To determine the effective cross-slope wavenumber, we employ complex empirical orthogonal function (CEOF) analysis of the low-frequency fluctuations. In a laterally bounded channel, the cross-slope structure may be represented by discrete normal modes with wavenumbers $\ell_n=n\pi/L$, where $n=1,2,3,\ldots$ is the cross-slope mode number and $L$ is the channel width \citep{pedlosky1987}. However, the disturbances observed here are concentrated over the continental slope and need not correspond to a single channel-wide normal mode. We therefore use CEOF analysis to identify the dominant coherent cross-slope structure of the spectral fluctuations \citep{horel1984,merrifield1990}. The analysis is performed separately over the upper- and lower-slope regions. The CEOF method retains both amplitude and spatial-phase information, allowing an effective cross-slope wavenumber to be estimated from the phase variation of the leading mode. Similar analyses have been used to investigate the spatial structure and propagation of observed TRWs \citep{chen2022}. The detailed procedure for calculating this wavenumber is described below, and the diagnosed values are reported in table~\ref{tab:ceof_summary}.

The remaining background parameters are determined separately for each Rossby number and slope region. The mean water-column depth and bottom slope gradient are determined from the prescribed slope, while the representative buoyancy frequency and background along-slope velocity are obtained from the DNS. This approach follows previous applications of linear TRW theory using representative local environmental conditions \citep{chen2022,guo2024seamount_trw}. The resulting background parameters are summarised in table~\ref{tab:trw_parameters}.

\begin{figure}
\centering
\includegraphics[width=\textwidth]{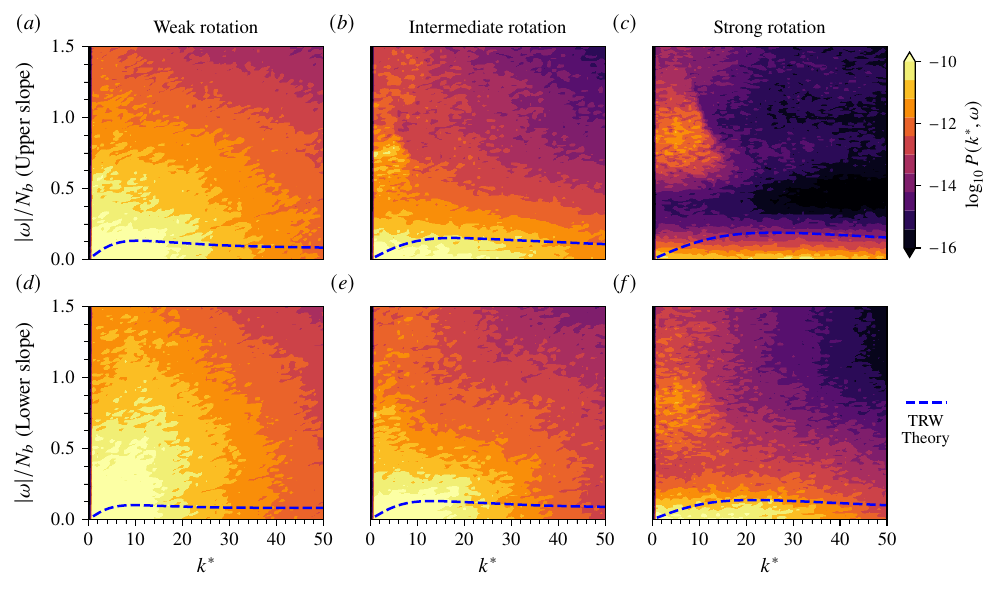}
\caption{
Doppler-corrected frequency--wavenumber spectra of the along-slope velocity fluctuations over the upper slope $\mathit{(a-c)}$ and lower slope $\mathit{(d-f)}$. Columns correspond to weak rotation ($Ro=1.4\times10^{-1}$), intermediate rotation ($Ro=3.6\times10^{-2}$), and strong rotation ($Ro=8.9\times10^{-3}$), respectively. Colours indicate $\log_{10}P(k^*,\omega)$ as a function of the dimensionless along-slope Fourier mode $k^*=kL_x/(2\pi)$ and the normalised intrinsic frequency $|\omega|/N_b$. The dashed curves represent the locally evaluated intrinsic linear TRW dispersion relation using regional background quantities diagnosed from the DNS and the corresponding effective cross-slope wavenumbers obtained from the leading low-frequency CEOF mode.
}
\label{fig:dispersion}
\end{figure}

Figure~\ref{fig:dispersion} shows the frequency--wavenumber distribution of the along-slope velocity fluctuations as rotation increases. At the largest Rossby number, figure~\ref{fig:dispersion}$\mathit{(a,d)}$, both slope regions exhibit a broad distribution of spectral energy extending towards relatively high frequencies. As $Ro$ decreases, the dominant spectral energy becomes progressively concentrated at lower frequencies, with substantial energy retained over a finite range of along-slope wavenumbers. This transition is particularly clear over both the upper and lower slopes. At $Ro=3.6\times10^{-2}$, the dominant spectral energy is concentrated mainly below $|\omega|/N_b\lesssim0.4$, while at $Ro=8.9\times10^{-3}$, it is concentrated mainly below $|\omega|/N_b\lesssim0.2$ in both slope regions. However, a separate higher-frequency energetic region remains visible over the upper slope at the smallest Rossby number.

In addition, the spectra are separated into two frequency ranges. The low-frequency component occupies $0.009\leq|\omega|/N_b\leq0.40$, whereas the higher-frequency component occupies $0.40<|\omega|/N_b\leq1.50$. The concentration of energy at low frequencies becomes particularly apparent at $Ro=3.6\times10^{-2}$ and $Ro=8.9\times10^{-3}$. In both the upper- and lower-slope regions, the low-frequency energy extends over a finite range of along-slope wavenumbers, whereas the higher-frequency fluctuations exhibit a broader frequency distribution. The threshold $|\omega|/N_b=0.40$ is therefore adopted to distinguish the dominant low-frequency variability from the higher-frequency fluctuations and is applied consistently to all cases.

The energetic low-frequency component is compared with the intrinsic linear topographic Rossby wave dispersion relation, shown by the dashed curves in figure~\ref{fig:dispersion}. For all three rotating cases, the theoretical dispersion curves lie within the identified low-frequency range and overlap regions of relatively high spectral power over both the upper and lower slopes. This comparison indicates that the observed low-frequency variability is qualitatively consistent with the predictions of linear topographic Rossby wave theory.

For each discrete along-slope Fourier mode $k^*=1,\ldots,50$, a representative frequency is identified from the spectral-power-weighted mean within the selected frequency band,
\begin{equation}
\omega_{\mathit{eff} }(k^*)
=
\frac{
\displaystyle\sum_{\omega\in\mathcal B}
\omega P(k^*,\omega)
}{
\displaystyle\sum_{\omega\in\mathcal B}
P(k^*,\omega)
},
\label{eq:omega_eff}
\end{equation}
where $\mathcal B$ denotes the selected low- or higher-frequency band of intrinsic frequencies. Thus, $\omega_{\mathit{eff}}(k^*)$ represents the spectral-power-weighted mean intrinsic frequency within the selected band for each discrete along-slope Fourier mode, calculated separately over the upper- and lower-slope regions.

The spectral-energy contributions of both frequency branches are evaluated separately over the upper- and lower-slope regions for $1\leq k^*\leq50$, with the results summarised in table~\ref{tab:ceof_summary}. The low-frequency component, $0.009\leq|\omega|/N_b\leq0.40$, accounts for $93.1$--$98.9\%$ of the spectral power over the upper slope and $95.2$--$98.4\%$ over the lower slope across the three rotating cases. The higher-frequency component, $0.40<|\omega|/N_b\leq1.50$, accounts for the remaining spectral power within the selected frequency ranges. Thus, the low-frequency component remains energetically dominant in both slope regions across all three rotating cases, although its relative contribution varies with Rossby number and cross-slope location. The subsequent CEOF analysis therefore focuses on the dominant low-frequency component to characterise its coherent cross-slope structure and assess its consistency with TRW dynamics.

For each discrete along-slope Fourier mode, the calculated $\omega_{\mathit{eff}}(k^*)$ within the low-frequency band defines the representative frequency of the corresponding spectral ridge. Because this frequency does not necessarily coincide with a discrete Fourier-frequency bin, the complex Fourier coefficients are extracted at the nearest resolved frequency, together with those at the immediately adjacent frequency bins, at each cross-slope sampling location. These ridge-associated coefficients retain both amplitude and phase information and are analysed using CEOF analysis to determine the dominant coherent cross-slope structure and its spatial phase variation. The phase variation of the leading CEOF mode is subsequently used to diagnose the effective cross-slope wavenumber required for the theoretical topographic Rossby-wave dispersion relation curve. The leading CEOF mode of the low-frequency component is expressed in amplitude--phase form \citep{horel1984} as
\begin{equation}
C_{1}(\hat y)
=
a_{1}(\hat y)
\exp\left[
\mathit{i}\phi_{1}(\hat y)
\right],
\label{eq:ceof_mode}
\end{equation}
where $a_{1}(\hat y)$ and $\phi_{1}(\hat y)$ denote the spatial amplitude and phase, respectively, of the leading CEOF mode in each slope region (upper and lower slope). The phase is unwrapped in the cross-slope direction and approximated by
\begin{equation}
\phi_{1}(\hat y)
\simeq
\phi_{0}
+
2\pi\ell_{ }^{*}\hat y,
\label{eq:ceof_phase_fit}
\end{equation}
where $\phi_{0}$ is the phase intercept and $\ell^{*}$ is the effective nondimensional cross-slope wavenumber. The effective cross-slope wavenumber is obtained from the gradient of the fitted phase, with the corresponding dimensional wavenumber and nondimensional wavelength given by
\begin{equation}
\ell^{*}
=\frac{1}{2\pi}\frac{\mathit{d}\phi_{1}}{\mathit{d}\hat y},
\qquad
\lambda^{*}
=\frac{1}{|\ell^{*}|}.
\label{eq:ell_nondim}
\end{equation}
where $\lambda_y^{*}$ is the corresponding nondimensional cross-slope wavelength. The diagnosed effective cross-slope wavenumbers are subsequently used in the regional linear TRW dispersion comparisons.
\begin{table}
\centering
\begin{tabular}{lcccccc}
\hline
$Ro$
& $E_{\mathit{low}}\;(\%)$
& $E_{\mathit{high}}\;(\%)$
& $E_{\mathit{CEOF1}}\;(\%)$
& $|\ell^*|$
& $\lambda^*$
& $R_\phi^2$ \\
\hline
\multicolumn{7}{l}{\textit{Upper slope}} \\
$1.42\times10^{-1}$
& 97.3 & 2.7 & 84.4 & 7.4 & 0.13 & 0.99 \\
$3.56\times10^{-2}$
& 98.9 & 1.1 & 85.5 & 12.7 & 0.08 & 0.99 \\
$8.89\times10^{-3}$
& 93.1 & 6.9 & 66.9 & 19.4 & 0.05 & 0.99 \\
\hline
\multicolumn{7}{l}{\textit{Lower slope}} \\
$1.42\times10^{-1}$
& 95.2 & 4.8 & 69.2 & 6.3 & 0.16 & 0.97 \\
$3.56\times10^{-2}$
& 97.9 & 2.1 & 71.8 & 9.2 & 0.11 & 0.98 \\
$8.89\times10^{-3}$
& 98.4 & 1.6 & 84.5 & 16.2 & 0.06 & 0.99 \\
\hline
\end{tabular}
\caption{
Summary of the spectral-energy partition and leading CEOF mode characteristics over the upper- and lower-slope regions. Here $E_{\mathit{low}}$ and $E_{\mathit{high}}$ denote the percentages of spectral power contained in the low- and higher-frequency bands, respectively, evaluated over $1\leq k^*\leq50$. The quantity $E_{\mathit{CEOF1}}$ represents the percentage of variance explained by the leading CEOF mode of the selected low-frequency ridge coefficients. The quantity $R_\phi^2$ denotes the coefficient of determination of the amplitude-weighted linear fit to the unwrapped phase of the leading CEOF mode.
}
\label{tab:ceof_summary}
\end{table}

Table~\ref{tab:ceof_summary} summarises the leading CEOF mode characteristics of the low-frequency component over the upper- and lower-slope regions. Over the upper slope, the leading CEOF mode accounts for $84.4\%$, $85.5\%$ and $66.9\%$ of the variance within the selected low-frequency ridge for $Ro=1.42\times10^{-1}$, $3.56\times10^{-2}$ and $8.89\times10^{-3}$, respectively. The corresponding contributions over the lower slope are $69.2\%$, $71.8\%$ and $84.5\%$. The leading-mode variance contribution is greater over the upper slope at the two larger Rossby numbers, whereas the lower slope exhibits a greater contribution at the smallest Rossby number. The unwrapped cross-slope phase is well approximated by a linear variation over the fitted intervals, with $R_\phi^2\geq0.97$ across all regional analyses. As $Ro$ decreases from $1.42\times10^{-1}$ to $8.89\times10^{-3}$, the magnitude of the effective nondimensional cross-slope wavenumber increases from $|\ell^*|=7.4$ to $19.4$ over the upper slope and from $6.3$ to $16.2$ over the lower slope. The corresponding nondimensional cross-slope
wavelength decreases from $0.13$ to $0.05$ over the upper slope and from $0.16$ to $0.06$ over the lower slope.

Overall, the spectral analysis identifies low-frequency fluctuations whose frequency--wavenumber characteristics are qualitatively consistent with the predictions of linear topographic Rossby wave theory.

\subsection{Phase velocity diagnosis}
\label{sec:phase_velocity_diagnisis}

To examine the along-slope propagation of the dominant low-frequency component identified in the spectral analyses presented in section~\ref{sec:trw}, the along-slope velocity fluctuations are reconstructed from selected Fourier components. The reconstruction is performed at representative cross-slope locations, $y/L=0.25$ and $0.40$, within the upper- and lower-slope regions, respectively. Only Fourier components satisfying $0.009\leq|\omega|/N_b\leq0.40$ and $1\leq k^*\leq15$ are retained. This wavenumber restriction isolates the larger-scale structures associated with the dominant low-frequency component.

\begin{figure}
\centering
\includegraphics[width=\textwidth]
{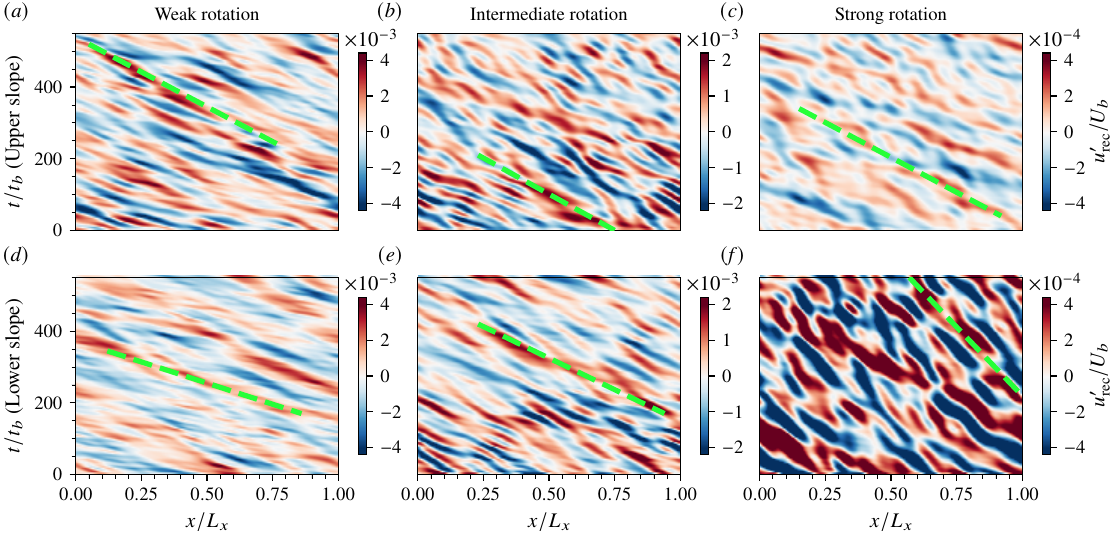}
\caption{
Hovm\"oller diagrams of the reconstructed low-frequency along-slope velocity fluctuations, $u'_{\mathit{rec}}/U_b$, at representative cross-slope locations within the upper-slope region ($y/L=0.25$, panels $\mathit{a, b, c}$) and lower-slope region ($y/L=0.40$, panels $\mathit{d, e,f}$). Here, $u'_{\mathit{rec}}$ denotes the along-slope velocity fluctuation reconstructed from the selected Fourier components. Columns correspond to $Ro=1.42\times10^{-1}$, $3.56\times10^{-2}$ and $8.89\times10^{-3}$, respectively. The reconstruction retains Fourier components satisfying $0.009\leq|\omega|/N_b\leq0.40$ and $1\leq k^*\leq15$. The green dashed lines indicate the inclination of phase bands, highlighting differences in along-slope phase propagation. Time is normalised by the buoyancy time scale, $t_b=L/U_b$. The colour scales differ between Rossby numbers to resolve the progressively weaker velocity fluctuations, while the same scale is used for both slope regions at each Rossby number.
}
\label{fig:hovmoller_lowk}
\end{figure}

Figure~\ref{fig:hovmoller_lowk} shows alternating positive and negative along-slope velocity anomalies forming coherent inclined bands in the $(x,t)$ plane at the sampled locations within both slope regions. The inclination of these bands indicates that the dominant low-frequency variability is associated with propagating structures in the along-slope direction. The green dashed lines indicate the inclination of representative bands, indicating differences in their propagation characteristics. Over the upper slope, figure~\ref{fig:hovmoller_lowk}$\mathit{(a,b,c)}$, coherent inclined bands are present for all three Rossby numbers, with no strong systematic change in their inclination as $Ro$ decreases. A clearer variation with rotation is observed over the lower slope, figure~\ref{fig:hovmoller_lowk}$\mathit{(d,e,f)}$, where the phase bands become progressively steeper as $Ro$ decreases from $1.42\times10^{-1}$ to $8.89\times10^{-3}$. This steepening indicates a reduction in the magnitude of the along-slope phase velocity with increasing rotation.

\begin{figure}
\centering
\includegraphics[width=\textwidth]
{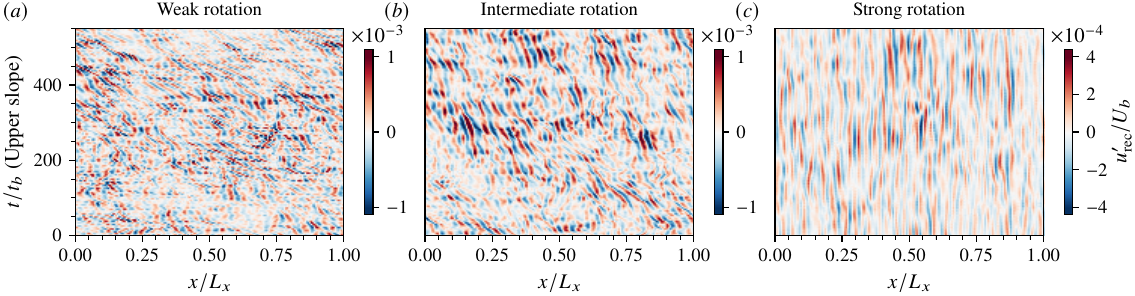}
\caption{
Hovm\"oller diagrams of the reconstructed higher-wavenumber ($15< k^*\leq50$) along-slope velocity fluctuations, $u'_{\mathit{rec}}/U_b$, at a representative cross-slope location ($y/L=0.25$) within the upper-slope region. Panels $\mathit{(a,b,c)}$ correspond to $Ro=1.42\times10^{-1}$, $3.56\times10^{-2}$ and $8.89\times10^{-3}$, respectively. Here, $u'_{\mathit{rec}}$ denotes the along-slope velocity fluctuation reconstructed from the selected Fourier components. The reconstruction retains the higher-wavenumber components without imposing a frequency-band restriction. Time is normalised by the buoyancy time scale, $t_b=L/U_b$.
}
\label{fig:hovmoller_highk}
\end{figure}

To examine the finer-scale contribution to the along-slope variability, figure~\ref{fig:hovmoller_highk} shows the corresponding upper-slope reconstruction for the higher-wavenumber range $15< k^*\leq50$, without imposing a frequency-band restriction. The resulting Hovm\"oller diagrams exhibit substantially shorter along-slope scales than those in figure~\ref{fig:hovmoller_lowk}. At $Ro=1.42\times10^{-1}$, inclined features remain visible, although they are finer and less spatially continuous than the low-wavenumber structures. As $Ro$ decreases, the bands become progressively steeper. At $Ro=3.56\times10^{-2}$, much of the variability consists of short, nearly vertical features, while at $Ro=8.89\times10^{-3}$ the reconstruction is dominated by steep, closely spaced bands with little horizontal displacement over the analysis interval. The corresponding lower-slope reconstruction exhibits similar behaviour, although these results are not shown.

To quantify the along-slope propagation characteristics identified in the Hovm\"oller diagrams in figure~\ref{fig:hovmoller_lowk}, the intrinsic phase velocity of the dominant low-frequency spectral component is calculated separately for the upper- and lower-slope regions using the spectral-power-weighted mean intrinsic frequency defined in (\ref{eq:omega_eff}). The analysis is restricted to the same low-frequency range $0.009\leq|\omega|/N_b\leq0.40$ and large-scale along-slope wavenumbers $1\leq k^*\leq15$. For each discrete along-slope wavenumber, indexed by $i$, the intrinsic phase velocity is calculated as
\begin{equation}
c_{p,i}
=
\frac{\omega_{\mathit{eff}}(k_i^*)}{k_i},
\end{equation}
where   $\omega_{\mathit{eff}}(k_i^*)$ is the corresponding energy-weighted intrinsic ridge frequency evaluated over the selected frequency branch. The intrinsic phase velocity for each slope region is then obtained by weighting the phase velocity at each retained wavenumber by its corresponding spectral power,
\begin{equation}
\overline{c}_p
=
\frac{
\displaystyle\sum_i E_i c_{p,i}
}{
\displaystyle\sum_i E_i
},
\label{eq:energy_weighted_phase_velocity}
\end{equation}
where $E_i=\sum_{\omega}P(k_i^*,\omega)$ is the spectral power summed over the selected low-frequency range at the $i$th along-slope wavenumber, calculated separately for the upper- and lower-slope regions. Negative phase velocities indicate propagation in the negative along-slope direction relative to the representative background along-slope flow.

\begin{figure}
\centering
\includegraphics[width=0.65\textwidth]
{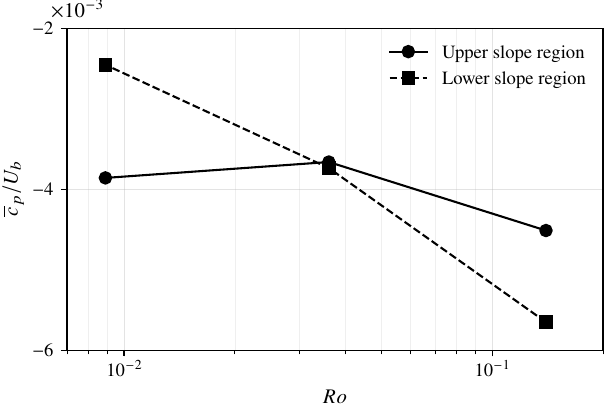}
\caption{
Normalised energy-weighted intrinsic phase velocity, $\overline{c}_p/U_b$, of the dominant low-frequency component as a function of Rossby number over the upper- and lower-slope regions. The solid line with circles represents the upper slope, while the dashed line with squares represents the lower slope. The phase velocity is calculated over the low-frequency range $0.009\leq|\omega|/N_b\leq0.40$ and the along-slope wavenumber range $1\leq k^*\leq15$. Negative values indicate propagation in the negative along-slope direction relative to the representative background along-slope flow.
}
\label{fig:phase_speed_low}
\end{figure}

Figure~\ref{fig:phase_speed_low} shows the normalised energy-weighted intrinsic phase velocity over both slope regions for the three rotating cases. Over the upper slope, $\overline{c}_p/U_b$ varies relatively weakly with Rossby number, with values of approximately $-4.5\times10^{-3}$, $-3.7\times10^{-3}$ and $-3.9\times10^{-3}$ as $Ro$ decreases from $1.42\times10^{-1}$ to $8.89\times10^{-3}$. Over the lower slope, the corresponding values are approximately $-5.7\times10^{-3}$, $-3.8\times10^{-3}$ and $-2.5\times10^{-3}$. The magnitude of the intrinsic phase velocity therefore decreases with increasing rotation over the lower slope, whereas the upper-slope values exhibit a comparatively weak dependence on rotation.

These trends are qualitatively consistent with the Hovm\"oller diagrams in figure~\ref{fig:hovmoller_lowk}, which show progressively steeper phase bands over the lower slope as rotation increases, while the upper-slope bands exhibit relatively little systematic change in inclination.

\section{Discussion}
\label{sec:discussion}

\subsection{Mean circulation, cross-slope transport and overturning}

The present results show that rotation and sloping topography reorganise the pathways of buoyancy-driven transport. In the non-rotating case, dense fluid formed at the upper boundary descends directly across the slope and generates a broad domain-scale overturning. With rotation, cross-slope and vertical motions are strongly suppressed, and the flow is redirected into oppositely directed along-slope currents. The formation of these currents is consistent with the geostrophic adjustment of dense flow over sloping topography \citep{smith1977experiments,condie1995descent}. The surface-intensified positive transport decreases approximately as $\psi_z^{+}\propto Ro^{0.59}$, whereas the magnitude of the negative, slope-attached transport follows the weaker dependence $|\psi_z^{-}|\propto Ro^{0.38}$. Both transport branches therefore weaken as $Ro$ decreases. These exponents are based on only three rotating simulations and can be interpreted as dependencies over the present parameter range rather than as general scaling laws.

The analysis of cold-water transport further shows how rotation influences cross-slope exchange. Colder water is transported downslope, while relatively warmer water is transported upslope, resulting in zero net time-averaged volume transport across each vertical section. The magnitude of this exchange decreases with increasing rotation, consistent with the suppression of the large-scale cross-slope circulation. The mean--fluctuation decomposition further shows that temporal fluctuations contribute substantially to the cumulative cold-water transport, highlighting the importance of distinguishing the mean circulation from the total cross-slope exchange of colder fluid.

The slope-attached mean current is concentrated within a bottom boundary layer whose thickness decreases with increasing rotation, approximately following the classical Ekman scaling. The diagnosed thickness scales as $\delta_E/L\propto Ro^{0.25}$ and $Ro^{0.26}$ at $y/L=0.2$ and $0.3$, respectively. The cold-water transport within this layer is dominated by the mean contribution at both sampling locations across all three rotating cases, while the fluctuation contribution remains negligible. Importantly, the bottom boundary layer carries only $0.8$--$4.1,\%$ of the net downslope cold-water transport across the two sampling sections for $\widehat{T}_c=0.006$, with the contribution decreasing as rotation increases. Thus, despite the presence of a mean current within the bottom boundary layer, most of the net downslope cold-water transport occurs outside this layer. These findings are qualitatively consistent with \citet{tanaka2006effects}, who demonstrated that although bottom Ekman transport contributes to dense-water descent, eddy-mediated transport becomes important following the development of baroclinic instability.

The residual overturning reveals distinct responses to rotation between the near-surface and interior circulations. The near-surface circulation weakens relatively slowly with decreasing Rossby number, following $\psi_{\mathit{surf}}/\psi_0\propto Ro^{0.30}$, whereas the interior overturning decreases more rapidly, following $\psi_{\mathit{int}}/\psi_0\propto Ro^{0.78}$. This contrast indicates that increasing rotation suppresses interior overturning more strongly than the near-surface circulation, progressively weakening the connection between surface buoyancy forcing and the deep interior. 
This behaviour is consistent with the increasing confinement of the time-mean cross-slope flow near the sloping boundary, although the temperature-conditioned transport analysis indicates that fluctuating motions continue to contribute to cold-water exchange outside the bottom boundary layer. At the same time, the upper thermal boundary layer thickens according to $\delta_T/L\propto Ro^{-0.21}$, consistent with reduced vertical advective exchange under stronger rotation. This behaviour agrees qualitatively with previous studies of rotating horizontal convection
\citep{barkan2013rotating,vreugdenhil2017geostrophic,
sohail2018convection,Vreugdenhil_Gayen_Griffiths_2019,
gayen2022rotating}.

These findings indicate that rotation reduces the strength of the buoyancy-driven circulation and modifies the pathways of cross-slope exchange. As rotation increases, the interior overturning weakens and the upper thermal boundary layer thickens, while a thinner bottom Ekman-like layer continues to support a mean downslope flow. However, this layer carries only a small fraction of the net downslope cold-water transport, with most of the transport occurring outside the bottom boundary layer, where temporal fluctuations contribute substantially to cross-slope cold-water exchange.

\subsection{Low-frequency variability and topographic Rossby-wave dynamics}

Rotation also influences the unsteady variability of the flow. At the largest Rossby number, the frequency--wavenumber spectra exhibit a relatively broad distribution of energy, whereas decreasing $Ro$ produces a stronger concentration of spectral energy at low frequencies and finite along-slope wavenumbers. Within the spectral ranges analysed, the low-frequency component, $0.009\leq|\omega|/N_b\leq0.40$, accounts for $93.1$--$98.9,\%$ of the spectral power over the upper slope and $95.2$--$98.4,\%$ over the lower slope. The higher-frequency component, $0.40<|\omega|/N_b\leq1.50$, contributes the remaining power within these selected ranges and is more broadly distributed in frequency. The low-frequency component therefore dominates the analysed variability in both slope regions.

The spectral and cross-slope structure analyses provide further evidence that the low-frequency variability is consistent with topographic Rossby wave dynamics. The locally evaluated linear topographic Rossby wave dispersion relation overlaps regions of relatively high spectral power within the low-frequency range over a finite range of along-slope wavenumbers. The complex empirical orthogonal function analysis further identifies a coherent cross-slope phase structure, with the leading mode accounting for a large proportion of the selected low-frequency ridge variance in both slope regions. As $Ro$ decreases, the magnitude of the effective nondimensional cross-slope wavenumber increases from $|\ell^*|=7.4$ to $19.4$ over the upper slope and from $6.3$ to $16.2$ over the lower slope. This indicates that the dominant coherent low-frequency structure develops a shorter cross-slope phase scale under stronger rotation. Furthermore, because the bottom slope is steep and spatially varying, and the effective cross-slope wavenumbers used in the theoretical curves are diagnosed from the DNS, the dispersion comparison can be interpreted as a consistency test rather than an independent quantitative validation of linear TRW theory.

The Hovm\"oller reconstructions further illustrate the along-slope propagation and spatial organisation of the velocity fluctuations. The low-frequency, low-wavenumber component exhibits coherent inclined structures over both slope regions, consistent with systematic along-slope propagation (figure~\ref{fig:hovmoller_lowk}). Over the upper slope, the inclination of these structures changes relatively low with rotation, whereas over the lower slope, the phase bands become significantly steeper as $Ro$ decreases, indicating a reduction in their along-slope propagation speed. In contrast, the higher-wavenumber reconstruction over the upper slope exhibits shorter, less spatially continuous structures, with increasingly steep, nearly vertical bands under stronger rotation (figure~\ref{fig:hovmoller_highk}). The energy-weighted intrinsic phase-velocity results provide a quantitative complement to the Hovm\"oller analysis and show a distinct regional dependence on rotation. Over the upper slope, the magnitude of the intrinsic phase velocity varies relatively weakly with Rossby number, whereas over the lower slope it decreases progressively with increasing rotation. These trends are qualitatively consistent with the Hovm\"oller observations. In both regions, the intrinsic phase velocity indicates propagation in the negative along-slope direction relative to the background flow.

Previous studies have shown that topographic Rossby waves can develop through baroclinic instability of dense overflows and interact with the descending dense current \citep{han2022trw}. \citet{han2023controls} further showed that downslope dense-water transport is controlled by bottom friction over steep slopes, enhanced by wave-associated form stress over intermediate slopes, and inhibited by coherent eddies over gentle slopes. Unlike these prescribed-overflow configurations, the dense fluid, background stratification, and along-slope currents in the present simulations develop together in response to continuous surface thermal forcing. The present results demonstrate that mean bottom boundary-layer transport, substantial fluctuating cross-slope cold-water transport and coherent topographic Rossby waves variability coexist within the same buoyancy-driven circulation. Future work could focus on quantifying the contribution of these wave motions to the fluctuating cross-slope transport.

\subsection{Relevance to the Antarctic continental margin}
\label{sec:antarctic_relevance}

The present configuration is designed to examine the fundamental dynamics of buoyancy-driven shelf--slope flow rather than represent a specific Antarctic region. Around Antarctica, dense water formed on the continental shelf crosses the shelf break and descends the continental slope, where rotation and topography favour along-slope geostrophic
flow and constrain cross-slope transport \citep{thompson2018antarctic}. The interaction between buoyancy-driven downslope motion and rotationally constrained along-slope flow is therefore relevant to the dynamics examined here. For comparison, \citet{sohail2018convection} estimated a characteristic Southern Ocean Rossby number of approximately $0.025$, comparable in magnitude to the intermediate-rotation case considered here. However, their estimate is based on domain-scale parameters, and differences in the characteristic scales and governing conditions should be considered when comparing the two systems.

The mean circulation exhibits several qualitative similarities with Antarctic continental-margin flows. Around much of Antarctica, the Antarctic Slope Current flows predominantly westward along the continental slope, with bottom-intensified currents developing in regions influenced by dense shelf-water outflows \citep{thompson2018antarctic}. The negative slope-attached current in the present simulations exhibits a similar near-bottom concentration of along-slope flow. The oppositely directed, surface-intensified positive current may also be qualitatively compared with the eastward-flowing Antarctic Circumpolar Current, although the latter is influenced by wind forcing, buoyancy gradients, eddies and interactions with topography, which are not fully represented in the present configuration. Furthermore, the mean downslope transport within the bottom boundary layer resembles an important aspect of Antarctic dense-water export, where dense shelf water descends the continental slope and mixes with surrounding waters, contributing to the formation of Antarctic Bottom Water \citep{baines1998observations,thompson2018antarctic}. These similarities highlight the role of rotation and sloping topography in organising along-slope circulation while permitting buoyancy-driven downslope transport.
 
The low-frequency variability provides a further connection with Antarctic continental-slope dynamics. Previous studies have shown that dense shelf-water outflows can generate topographic Rossby waves, with their characteristics influenced by bottom slope, stratification, density differences and background along-slope currents \citep{marques2014topographic,han2022trw,han2023controls}. In the present simulations, the dense current, background stratification, and along-slope flow develop together under continuous surface buoyancy forcing, rather than through a prescribed dense-water overflow. The emergence of coherent propagating low-frequency variability consistent with topographic Rossby waves dynamics indicates that similar topographically controlled motions can develop within a surface-forced shelf--slope circulation. This provides a qualitative connection with the low-frequency variability associated with Antarctic dense-water outflows, although quantifying the influence of these waves on downslope dense-water transport is beyond the scope of the present study.

The present simulations differ from Antarctic continental-margin conditions in several respects. The model uses smooth idealised topography and thermal forcing alone, whereas Antarctic dense-water formation involves both temperature and salinity, with additional influences from wind, tides, ice shelves and complex bathymetry. In particular, submarine canyons can facilitate intrusions of warm offshore water onto the continental shelf \citep{ong2024intrusions}. Furthermore, oceanic Rayleigh numbers are substantially larger than those considered here, and the Rossby-number dependencies are based on only three rotating cases. These differences should be considered when extending the present findings to oceanic conditions. Future work could examine how more realistic topography, including canyons, influences the circulation and cross-slope transport.

\section{Conclusions}
\label{sec:conclusions}

This study used direct numerical simulations of rotating horizontal convection over an idealised continental shelf--slope system to examine how rotation modifies buoyancy-driven circulation, downslope transport and the associated low-frequency variability. A non-rotating case and three rotating cases spanning Rossby number $Ro=1.4\times10^{-1}$ to $8.9\times10^{-3}$ were considered to investigate the transition from a broad gravity-driven overturning circulation to a rotationally constrained slope flow.

Unlike the previous studies in which dense water is supplied through a prescribed overflow, the present simulations allow dense-water formation, the resulting circulation and low-frequency variability to develop together under continuous surface buoyancy forcing. This allows the effects of rotation on the mean circulation and associated unsteady motions over sloping topography to be examined within the same system.

Rotation reorganises the mean circulation. The broad downslope flow and domain-scale overturning of the non-rotating case are replaced by oppositely directed slope-attached and surface-intensified along-slope currents, while interior cross-slope and vertical motions are strongly suppressed. Both along-slope transport branches weaken with increasing rotation, with the surface-intensified positive branch exhibiting a stronger Rossby-number dependence than the negative slope-attached branch.

Rotation also weakens the residual overturning, which exhibits distinct interior and near-surface cells, with the interior cell weakening more rapidly as rotation increases. Meanwhile, the upper thermal boundary layer thickens, whereas the bottom Ekman boundary layer becomes thinner, with transport within this layer generally dominated by the mean flow. However, the bottom Ekman boundary layer accounts for only a small fraction of the net downslope dense-water transport at the sampled sections for the selected temperature threshold. In contrast, temporal fluctuations make a significant contribution to downslope dense-water transport across the full water column. These findings indicate that, despite the rotational suppression of large-scale overturning, downslope dense-water transport extends beyond the bottom Ekman boundary layer and is largely driven by temporal fluctuations.

Rotation further modifies the low-frequency variability over the continental slope. The frequency--wavenumber spectra reveal that the flow is dominated by low-frequency variability, with most of the spectral power concentrated at low frequencies within the analysed spectral ranges. The locally evaluated linear topographic Rossby wave dispersion relation overlaps regions of high spectral power within the low-frequency range over a finite range of along-slope wavenumbers. The Hovm\"oller reconstructions alos show systematic along-slope propagation of these fluctuations. The magnitude of the intrinsic phase velocity varies weakly with rotation over the upper slope but decreases with increasing rotation over the lower slope. These characteristics support the interpretation that the observed low-frequency variability is consistent with topographic Rossby-wave dynamics.

Future studies could further investigate the role of these wave motions in downslope dense-water transport and incorporate more realistic topography, additional forcing processes and a wider range of governing parameters to assess the relevance of these dynamics to Antarctic continental-margin flows.

\section{Acknowledgments}

This research was supported by the Australian Research Council (ARC) Special Research Initiative, Australian Centre for Excellence in Antarctic Science (Project Number SR200100008) and by the ARC Discovery Project (DP240102823).  C.V. was supported by ARC DECRA Fellowship Grant (DE220101027) and S.I. was supported by the Melbourne Research Scholarship. This project received grant funding from the Australian Government, and contributes to delivering the Australian Antarctic Science Decadal Strategy. Simulations were undertaken using the National Computational Infrastructure in Canberra and Pawsey Supercomputing Centre in Perth, Australia.  

\section{Declaration of Interests}
The authors report no conflict of interest.

\begin{appen}

\appendix
\section{Energy-budget and spatial-resolution assessment}
\label{app:Energy_budget}

\subsection{Global energy budget}

The time evolution of the global kinetic-energy budget is examined to assess the approach of the numerical solutions towards statistically stationary conditions. The budget terms considered are the buoyancy conversion ($B$), viscous dissipation ($\varepsilon$) and kinetic-energy tendency ($\mathit{d}KE/\mathit{d}t$). Each term is normalised by the time-averaged rate of conversion of internal energy to potential energy, $\overline{\Phi_i}$, evaluated separately for each case. Here, an overbar denotes a time average, while $\langle\cdot\rangle_V$ denotes integration over the domain volume. Under statistically stationary conditions, the time-averaged and volume-integrated kinetic-energy tendency becomes negligible, such that $\langle\overline{\mathit{d}KE/\mathit{d}t}\rangle_V\approx 0$, and the corresponding buoyancy conversion approximately balances viscous dissipation, $\langle\overline{B}\rangle_V\approx\langle\overline{\varepsilon}\rangle_V$. Similarly, the potential-energy budget requires $\langle\overline{B}\rangle_V\approx\overline{\Phi_i}$ when the potential-energy tendency is negligible and the time-averaged net surface heat input vanishes under thermally stationary conditions \citep{Winters1995,Hughes2009}. 

Figure~\ref{fig:ke_budget_evolution} shows the evolution of the normalised kinetic-energy-budget terms for the non-rotating and three rotating cases. The initial adjustment is characterised by relatively large buoyancy-conversion and viscous dissipation terms, whose magnitudes gradually decrease with time. Transient fluctuations are also observed in the rotating cases before the flow approaches its late-time regime. The corresponding time-averaged quantities are summarised in table~\ref{tab:ke_budget}. The magnitudes of the normalised buoyancy conversion, $|\langle\overline{B}\rangle_V|/\overline{\Phi_i}$, and viscous dissipation, $\langle\overline{\varepsilon}\rangle_V/\overline{\Phi_i}$, remain of order unity across all rotating cases. The relatively small normalised time-averaged kinetic-energy
tendencies, $|\langle\overline{\mathit{d}KE/\mathit{d}t}\rangle_V|/ \overline{\Phi_i}$, indicate that the rotating cases have approached statistically stationary conditions over the selected averaging intervals, with buoyancy conversion approximately balancing viscous dissipation. The non-rotating case is considered to have reached a quasi-steady state, as computational resources were prioritised for the rotating cases, which are the primary focus of this study.

Due to the high computational cost of DNS, the weak- and intermediate-rotation cases were initially simulated at a relatively coarse spatial resolution. The resulting flow fields were subsequently interpolated onto finer computational grids at the target DNS resolution. Following interpolation, the simulations were continued at the final DNS resolution for a sufficiently long period to allow the adjustment transients to decay and statistically stationary conditions to be established. The early transient fluctuations observed in figures~\ref{fig:ke_budget_evolution}$\mathit{(b,c)}$ occur during the initial adjustment and grid-refinement stages. As the present study focuses on the statistically stationary regime, all subsequent analyses and calculations were performed using high-resolution DNS data
obtained after statistical stationarity had been established.

\begin{figure}
\centering
\includegraphics[width=1.0\textwidth]{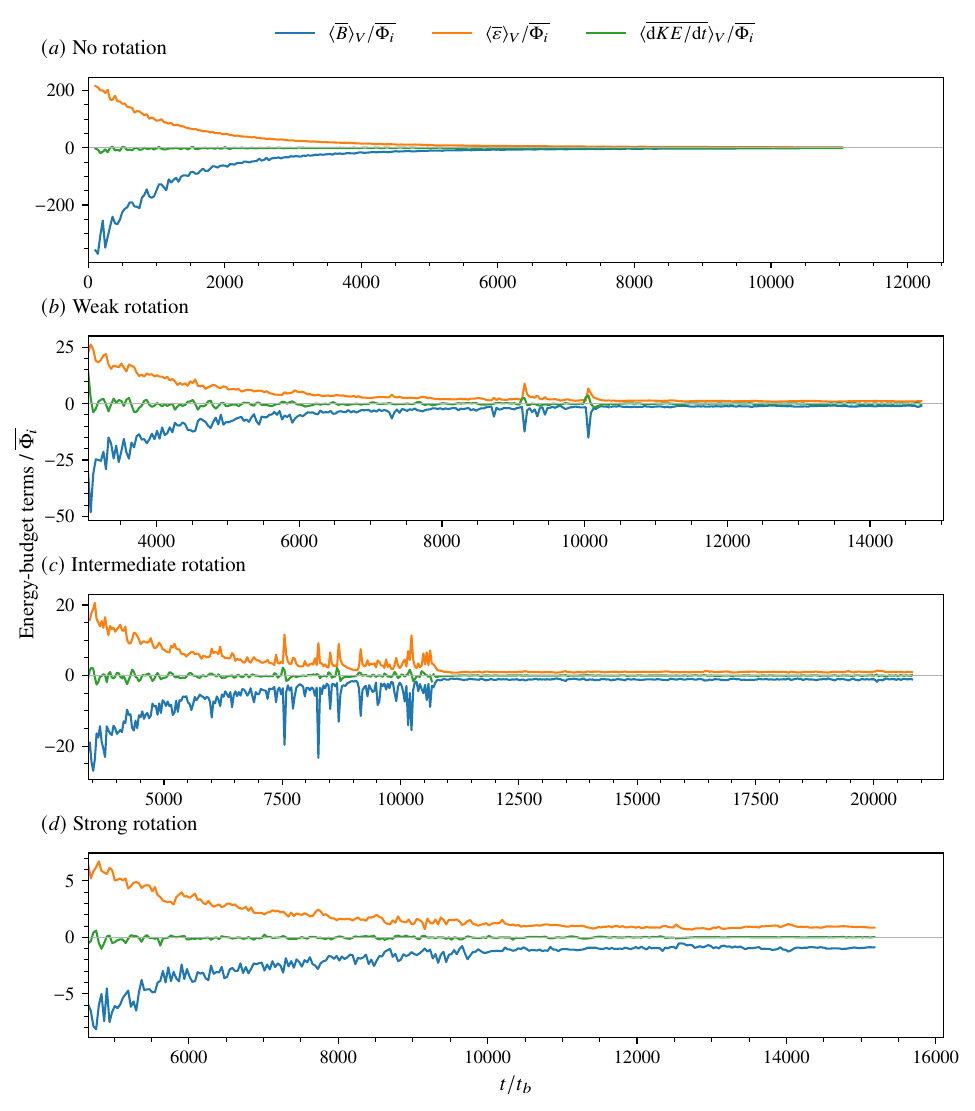}
\caption{
Time evolution of the energy budget terms for (a) the non-rotating case ($Ro=\infty$), (b) weak rotation ($Ro=1.4\times10^{-1}$), (c) intermediate rotation ($Ro=3.6\times10^{-2}$) and (d) strong rotation ($Ro=8.9\times10^{-3}$). The terms shown are the buoyancy conversion ($-B$), viscous dissipation ($\varepsilon$) and kinetic-energy tendency ($\mathit{d}KE/\mathit{d}t$), each normalised by the time-averaged conversion rate of internal energy to potential energy ($\langle\Phi_i\rangle$). Time is normalised by the buoyancy time scale $t_b=L/U_b$.
}
\label{fig:ke_budget_evolution}
\end{figure}

\begin{table}
\centering
\small
\begin{tabular}{lcccc}
\hline
Case & $Ro$
& $|\langle\overline{B}\rangle_V/\overline{\Phi_i}|$
& $\langle\overline{\varepsilon}\rangle_V/\overline{\Phi_i}$
& $|\langle\overline{\mathit{d}KE/\mathit{d}t}\rangle_V/\overline{\Phi_i}|$ \\[3pt]
\hline
Non-rotating
& $\infty$
& $1.3554$
& $1.3605$
& $0.04683$ \\

Weak rotation
& $1.4\times10^{-1}$
& $1.0697$
& $1.0318$
& $0.00884$ \\

Intermediate rotation
& $3.6\times10^{-2}$
& $1.0727$
& $1.0373$
& $0.00645$ \\

Strong rotation
& $8.9\times10^{-3}$
& $1.0051$
& $0.9638$
& $0.00397$ \\
\hline
\end{tabular}
\caption{
Time evolution of the normalised kinetic-energy-budget terms for $\mathit{(a)}$ the non-rotating case ($Ro=\infty$), $\mathit{(b)}$ weak rotation ($Ro=1.4\times10^{-1}$), $\mathit{(c)}$ intermediate rotation ($Ro=3.6\times10^{-2}$) and $\mathit{(d)}$ strong rotation ($Ro=8.9\times10^{-3}$). The terms shown are the buoyancy conversion ($B$), viscous dissipation ($\varepsilon$) and kinetic-energy tendency ($\mathit{d}KE/\mathit{d}t$), each normalised by the time-averaged rate of conversion of internal energy to potential energy ($\overline{\Phi_i}$). The early fluctuations in the weak- and intermediate-rotation cases occur during the initial adjustment and grid-refinement stages. Following interpolation onto the finer grids, the simulations were integrated further at the final DNS resolution until statistically stationary conditions were approached. Only the late-time, statistically stationary portions of the DNS-resolution simulations were used for the time-averaged diagnostics.
}
\label{tab:ke_budget}
\end{table}
\subsection{Spatial-resolution assessment}

The spatial resolution is further assessed for all simulated cases by comparing the local grid spacing with the Batchelor microscale, following the approach of \citet{gayen2014stability}. The Batchelor microscale characterises the small scale thermal fluctuations. For $Pr=5$, the Batchelor scale is smaller than the Kolmogorov scale and therefore provides a more restrictive measure of the spatial resolution required for the temperature field. The local Batchelor scale is calculated as
\begin{equation}
\eta_b
=
\frac{1}{\sqrt{Pr}}
\left(\frac{\nu^3}{\varepsilon_{\mathit{loc}}}\right)^{1/4}
=
\frac{\eta_K}{\sqrt{Pr}},
\end{equation}
where $\eta_K=(\nu^3/\varepsilon_{\mathit{loc}})^{1/4}$ is the Kolmogorov microscale and $\varepsilon_{\mathit{loc}}$ is the local viscous
dissipation rate. The ratios $\Delta_x/\eta_b$, $\Delta_y/\eta_b$ and $\Delta_z/\eta_b$ are evaluated using the local grid spacing in each coordinate direction.

This spatial-resolution assessment was performed for all simulated cases using data from the statistically stationary regime of the fully resolved DNS. For illustration, figure~\ref{fig:batchelor_resolution} presents the along-slope-averaged distributions of these ratios for the strongly rotating case ($Ro=8.9\times10^{-3}$). The same assessment was carried out for all other cases, but only the strongly rotating case is shown to illustrate the spatial variation of grid resolution relative to the Batchelor microscale.

\begin{figure}
\centering
\includegraphics[width=1.0\textwidth]{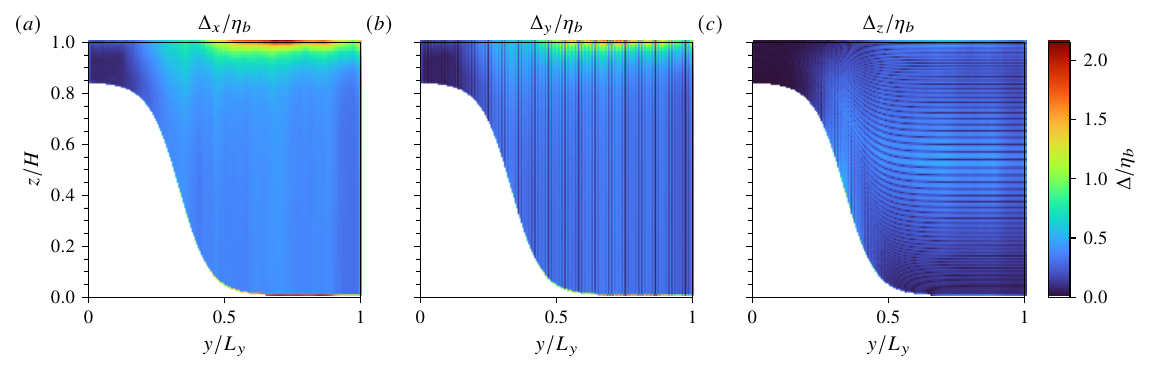}
\caption{
Along-slope-averaged distributions of the local grid-spacing ratios relative to the Batchelor microscale ($\eta_b$) for the strongly rotating case ($Ro=8.9\times10^{-3}$): $\mathit{(a)}$ $\Delta_x/\eta_b$, $\mathit{(b)}$ $\Delta_y/\eta_b$ and $\mathit{(c)}$$\Delta_z/\eta_b$. The averaging is performed along the periodic $x$-direction, and the resulting distributions are presented in the $y$--$z$ plane.
}
\label{fig:batchelor_resolution}
\end{figure}

\section{Slope-following coordinate transformation}
\label{app:slope_coordinates}

The local slope-parallel and bottom-normal unit vectors, denoted by $\hat{\boldsymbol e}_s$ and $\hat{\boldsymbol e}_n$, respectively, are defined relative to the Cartesian unit vectors $(\hat{\boldsymbol e}_y,\hat{\boldsymbol e}_z)$ at the reference locations $y/L=0.2$ and $0.3$. Here, $y$ denotes the fixed reference location, and $z_b(y)$ is the corresponding bottom elevation. The local bottom gradient is $s_b=\mathit{d}z_b/\mathit{d}y$, and the inclination angle is $\theta=\tan^{-1}(-s_b)$, evaluated at each reference location. The corresponding unit vectors are

\begin{equation}
\begin{aligned}
\hat{\boldsymbol e}_s
&=
\cos\theta\,\hat{\boldsymbol e}_y
-\sin\theta\,\hat{\boldsymbol e}_z,\\
\hat{\boldsymbol e}_n
&=
\sin\theta\,\hat{\boldsymbol e}_y
+\cos\theta\,\hat{\boldsymbol e}_z.
\end{aligned}
\label{eq:local_coordinate_transformation}
\end{equation}
Here, $\hat{\boldsymbol e}_s$ points downslope, while $\hat{\boldsymbol e}_n$ points into the fluid normal to the local bottom tangent, as illustrated in figure~\ref{fig:ekman_layer_geometry}(a). The corresponding slope-parallel velocity is
\begin{equation}
v^*=v\cos\theta-w\sin\theta,
\label{eq:slope_parallel_velocity}
\end{equation}
where $v$ and $w$ are the cross-slope and vertical velocity components, respectively.

The bottom-normal distance $n^*$ is measured from the local bottom reference point $(y,z_b(y))$ along $\hat{\boldsymbol e}_n$. The velocity profile used to diagnose the bottom boundary-layer thickness is evaluated along this direction, for which
\begin{equation}
y'=y+n^*\sin\theta,
\qquad
z=z_b(y)+n^*\cos\theta,
\label{eq:bottom_normal_coordinates}
\end{equation}
where $y'$ denotes the cross-slope coordinate along the local bottom-normal profile. Thus, the time- and along-slope-averaged slope-parallel velocity profile is
\begin{equation}
\left\langle\overline{v^*}\right\rangle_x(n^*)
=
\left\langle\overline{v^*}\right\rangle_x
\left(
y+n^*\sin\theta,\,
z_b(y)+n^*\cos\theta
\right).
\label{eq:bottom_normal_velocity_profile}
\end{equation}
This profile is used to determine the bottom boundary-layer thickness $\delta_E$ along the local bottom-normal direction at $y/L=0.2$ and $0.3$.
\end{appen}


\bibliographystyle{jfm}
\bibliography{jfm}

@article{whitehead1990experimental,
  author = {Whitehead, John A and Stern, Melvin E and Flierl, Glenn R and Klinger, Barry A},
  title = {Experimental observations of baroclinic eddies on a sloping bottom},
  journal = {Journal of Geophysical Research: Oceans},
  volume = {95},
  number = {C6},
  pages = {9585--9610},
  year = {1990},
  doi = {10.1029/JC095iC06p09585},
}

@article{mory1987coherent,
  author = {Mory, M and Stern, ME and Griffiths, RW},
  title = {Coherent baroclinic eddies on a sloping bottom},
  journal = {Journal of Fluid Mechanics},
  volume = {183},
  pages = {45--62},
  year = {1987},
}

@article{rossby1965thermal,
  author = {Rossby, H. T.},
  title = {On thermal convection driven by non-uniform heating from below: an experimental study},
  journal = {Deep Sea Research and Oceanographic Abstracts},
  volume = {12},
  number = {1},
  pages = {9--16},
  year = {1965},
  doi = {10.1016/0011-7471(65)91336-7},
}

@article{mullarney2004convection,
  author = {Mullarney, Julia C and Griffiths, Ross W and Hughes, Graham O},
  title = {Convection driven by differential heating at a horizontal boundary},
  journal = {Journal of Fluid Mechanics},
  volume = {516},
  pages = {181--209},
  year = {2004},
}

@article{annurev:/content/journals/10.1146/annurev.fluid.40.111406.102148,
  author = {Hughes, Graham O. and Griffiths, Ross W.},
  title = {Horizontal Convection},
  journal = {Annual Review of Fluid Mechanics},
  volume = {40},
  pages = {185--208},
  year = {2008},
  doi = {10.1146/annurev.fluid.40.111406.102148},
}

@article{gayen2022rotating,
  author = {Gayen, Bishakhdatta and Griffiths, Ross W.},
  title = {Rotating horizontal convection},
  journal = {Annual Review of Fluid Mechanics},
  volume = {54},
  pages = {105--132},
  year = {2022},
  doi = {10.1146/annurev-fluid-030121-115729},
}

@article{griffiths2013horizontal,
  author = {Griffiths, Ross W and Hughes, Graham O and Gayen, Bishakhdatta},
  title = {Horizontal convection dynamics: insights from transient adjustment},
  journal = {Journal of Fluid Mechanics},
  volume = {726},
  pages = {559--595},
  year = {2013},
}

@article{gayen2013energetics,
  author = {Gayen, Bishakhdatta and Griffiths, Ross W and Hughes, Graham O and Saenz, Juan A},
  title = {Energetics of horizontal convection},
  journal = {Journal of Fluid Mechanics},
  volume = {716},
  pages = {R10},
  year = {2013},
}

@article{Hignett_Ibbetson_Killworth_1981,
  author = {Hignett, Phillip and Ibbetson, Alan and Killworth, Peter D.},
  title = {On rotating thermal convection driven by non-uniform heating from below},
  journal = {Journal of Fluid Mechanics},
  volume = {109},
  pages = {161--187},
  year = {1981},
  doi = {10.1017/S0022112081000992},
}

@article{smith1977experiments,
  author = {Smith, Peter C},
  title = {Experiments with viscous source flows in rotating systems},
  journal = {Dynamics of Atmospheres and Oceans},
  volume = {1},
  number = {3},
  pages = {241--272},
  year = {1977},
}

@article{Griffiths_Killworth_Stern_1982,
  author = {Griffiths, R. W. and Killworth, Peter D. and Stern, Melvin E.},
  title = {Ageostrophic instability of ocean currents},
  journal = {Journal of Fluid Mechanics},
  volume = {117},
  pages = {343--377},
  year = {1982},
  doi = {10.1017/S0022112082001669},
}

@article{killworth1977mixing,
  author = {Killworth, Peter D},
  title = {Mixing of the {Weddell Sea} continental slope},
  journal = {Deep Sea Research},
  volume = {24},
  number = {5},
  pages = {427--448},
  year = {1977},
}

@incollection{nagata1993laboratory,
  author = {Nagata, Yutaka and Kimura, Ryuji and Honji, Hiroyuki and Yamazaki, Yasuhiro and Kawaguchi, Kazuhiro and Hosoyamada, Tokuzo},
  title = {Laboratory experiments of dense water descending on continental slope},
  booktitle = {{Deep Ocean} Circulation: Physical and Chemical Aspects},
  editor = {Teramoto, T.},
  series = {Elsevier Oceanography Series},
  volume = {59},
  pages = {333--350},
  year = {1993},
  publisher = {Elsevier},
}

@article{condie1995descent,
  author = {Condie, Scott A},
  title = {Descent of dense water masses along continental slopes},
  journal = {Journal of Marine Research},
  volume = {53},
  number = {6},
  pages = {897--928},
  year = {1995},
  doi = {10.1357/0022240953212936},
}

@article{lane1998eddy,
  author = {Lane-Serff, Gregory F and Baines, Peter G},
  title = {Eddy formation by dense flows on slopes in a rotating fluid},
  journal = {Journal of Fluid Mechanics},
  volume = {363},
  pages = {229--252},
  year = {1998},
}

@article{tanaka2006effects,
  author = {Tanaka, Kiyoshi},
  title = {Effects of the {Earth}'s rotation and bottom slope on a density current descending a sloping bottom},
  journal = {Journal of Geophysical Research: Oceans},
  volume = {111},
  number = {C11},
  pages = {C11018},
  year = {2006},
  doi = {10.1029/2006JC003677},
}

@article{sutherland2004rotating,
  author = {Sutherland, Bruce R and Nault, Joshua and Yewchuk, Kerianne and Swaters, Gordon E},
  title = {Rotating dense currents on a slope. Part 1. Stability},
  journal = {Journal of Fluid Mechanics},
  volume = {508},
  pages = {241--264},
  year = {2004},
}

@article{choboter2000baroclinic,
  author = {Choboter, Paul F and Swaters, Gordon E},
  title = {On the baroclinic instability of axisymmetric rotating gravity currents with bottom slope},
  journal = {Journal of Fluid Mechanics},
  volume = {408},
  pages = {149--177},
  year = {2000},
}

@article{Vreugdenhil_Gayen_Griffiths_2019,
  author = {Vreugdenhil, Catherine A. and Gayen, Bishakhdatta and Griffiths, Ross W.},
  title = {Transport by deep convection in basin-scale geostrophic circulation: turbulence-resolving simulations},
  journal = {Journal of Fluid Mechanics},
  volume = {865},
  pages = {681--719},
  year = {2019},
  doi = {10.1017/jfm.2019.64},
}

@article{nof1983translation,
  author = {Nof, Doron},
  title = {The translation of isolated cold eddies on a sloping bottom},
  journal = {Deep Sea Research Part A. Oceanographic Research Papers},
  volume = {30},
  number = {2},
  pages = {171--182},
  year = {1983},
}

@incollection{baines1998observations,
  author = {Baines, Peter G and Condie, Scott},
  title = {Observations and modelling of {Antarctic} downslope flows: A review},
  booktitle = {Ocean, Ice, and Atmosphere: Interactions at the {Antarctic} Continental Margin},
  editor = {Jacobs, Stanley S. and Weiss, Ray F.},
  series = {Antarctic Research Series},
  volume = {75},
  pages = {29--49},
  year = {1998},
  publisher = {American Geophysical Union},
  doi = {10.1029/AR075p0029},
}

@article{ADenseCurrentFlowingdownaSlopingBottominaRotatingFluid,
  author = {C. Cenedese and J. A. Whitehead and T. A. Ascarelli and M. Ohiwa},
  title = {A Dense Current Flowing down a Sloping Bottom in a Rotating Fluid},
  journal = {Journal of Physical Oceanography},
  volume = {34},
  number = {1},
  pages = {188--203},
  year = {2004},
  doi = {10.1175/1520-0485(2004)034<0188:ADCFDA>2.0.CO;2},
}

@article{vreugdenhil2017geostrophic,
  author = {Vreugdenhil, Catherine A and Griffiths, Ross W and Gayen, Bishakhdatta},
  title = {Geostrophic and chimney regimes in rotating horizontal convection with imposed heat flux},
  journal = {Journal of Fluid Mechanics},
  volume = {823},
  pages = {57--99},
  year = {2017},
}

@article{stewart2013connecting,
  author = {Stewart, Andrew L and Thompson, Andrew F},
  title = {Connecting {Antarctic} cross-slope exchange with {Southern Ocean} overturning},
  journal = {Journal of Physical Oceanography},
  volume = {43},
  number = {7},
  pages = {1453--1471},
  year = {2013},
}

@article{stewart2015eddy,
  author = {Stewart, Andrew L and Thompson, Andrew F},
  title = {Eddy-mediated transport of warm circumpolar deep water across the {Antarctic} shelf break},
  journal = {Geophysical Research Letters},
  volume = {42},
  number = {2},
  pages = {432--440},
  year = {2015},
}

@article{stewart2016eddy,
  author = {Stewart, Andrew L and Thompson, Andrew F},
  title = {Eddy generation and jet formation via dense water outflows across the {Antarctic} continental slope},
  journal = {Journal of Physical Oceanography},
  volume = {46},
  number = {12},
  pages = {3729--3750},
  year = {2016},
}

@article{nurser2004isopycnal,
  author = {Nurser, A. J. George and Lee, Mei-Man},
  title = {Isopycnal averaging at constant height. Part I: The formulation and a case study},
  journal = {Journal of Physical Oceanography},
  volume = {34},
  number = {12},
  pages = {2721--2739},
  year = {2004},
  doi = {10.1175/JPO2649.1},
}

@article{sohail2018convection,
  author = {Sohail, Taimoor and Gayen, Bishakhdatta and Hogg, Andrew McC},
  title = {Convection enhances mixing in the {Southern Ocean}},
  journal = {Geophysical Research Letters},
  volume = {45},
  number = {9},
  pages = {4198--4207},
  year = {2018},
}

@article{barkan2013rotating,
  author = {Barkan, Roy and Winters, Kraig B and Smith, Stefan G Llewellyn},
  title = {Rotating horizontal convection},
  journal = {Journal of Fluid Mechanics},
  volume = {723},
  pages = {556--586},
  year = {2013},
}

@article{gayen2014stability,
  author = {Gayen, Bishakhdatta and Griffiths, Ross W and Hughes, Graham O},
  title = {Stability transitions and turbulence in horizontal convection},
  journal = {Journal of Fluid Mechanics},
  volume = {751},
  pages = {698--724},
  year = {2014},
}

@misc{fischer2008nek5000,
  author = {Fischer, Paul F. and Lottes, James W. and Kerkemeier, Stefan G.},
  title = {{Nek5000}: Open-source spectral element {CFD} solver},
  year = {2008},
  howpublished = {\url{https://nek5000.mcs.anl.gov/}},
}

@book{pedlosky1987,
  author = {Pedlosky, Joseph},
  title = {Geophysical Fluid Dynamics},
  edition = {2},
  publisher = {Springer-Verlag},
  address = {New York},
  year = {1987},
  doi = {10.1007/978-1-4612-4650-3},
}

@article{horel1984,
  author = {Horel, J. D.},
  title = {Complex Principal Component Analysis: Theory and Examples},
  journal = {Journal of Climate and Applied Meteorology},
  volume = {23},
  number = {12},
  pages = {1660--1673},
  year = {1984},
  doi = {10.1175/1520-0450(1984)023<1660:CPCATA>2.0.CO;2},
}

@article{chen2022,
  author = {Chen, Juntian and Zhu, Xiao-Hua and Zheng, Hua and
             Nakamura, Hirohiko and Zhao, Ruixiang and Wang, Min and
             Park, Jae-Hun and Nishina, Ayako},
  title = {Observation of Topographic {Rossby} Waves Triggered by
             {Kuroshio} Path Meander in the {East China Sea}},
  journal = {Journal of Geophysical Research: Oceans},
  volume = {127},
  number = {8},
  pages = {e2022JC018667},
  year = {2022},
  doi = {10.1029/2022JC018667},
}

@article{merrifield1990,
  author = {Merrifield, M. A. and Guza, R. T.},
  title = {Detecting Propagating Signals with Complex Empirical
             Orthogonal Functions: A Cautionary Note},
  journal = {Journal of Physical Oceanography},
  volume = {20},
  number = {10},
  pages = {1628--1633},
  year = {1990},
  doi = {10.1175/1520-0485(1990)020<1628:DPSWCE>2.0.CO;2},
}

@article{ulloa2022development,
  author = {Ulloa, Hugo N. and Ram{\'o}n, Cintia L. and Doda, Tomy
             and W{\"u}est, Alfred and Bouffard, Damien},
  title = {Development of overturning circulation in sloping
             waterbodies due to surface cooling},
  journal = {Journal of Fluid Mechanics},
  volume = {930},
  pages = {A18},
  year = {2022},
  doi = {10.1017/jfm.2021.883},
}

@article{legg2009improving,
  author = {Legg, Sonya and Briegleb, Bruce and Chang, Yeon
             and Chassignet, Eric P. and Danabasoglu, Gokhan
             and Ezer, Tal and Gordon, Arnold L. and Griffies, Stephen
             and Hallberg, Robert and Jackson, Laura and Large, William
             and Ozgokmen, Tamay M. and Peters, Hartmut
             and Price, Jim and Riemenschneider, Ulrike and Wu, Wanli
             and Xu, Xiaobiao and Yang, Jiayan},
  title = {Improving Oceanic Overflow Representation in Climate Models:
             The Gravity Current Entrainment Climate Process Team},
  journal = {Bulletin of the American Meteorological Society},
  volume = {90},
  number = {5},
  pages = {657--670},
  year = {2009},
  doi = {10.1175/2008BAMS2667.1},
}

@article{pedlosky1967laboratory,
  author = {Pedlosky, J. and Greenspan, H. P.},
  title = {A simple laboratory model for the oceanic circulation},
  journal = {Journal of Fluid Mechanics},
  volume = {27},
  number = {2},
  pages = {291--304},
  year = {1967},
  doi = {10.1017/S0022112067000321},
}

@article{rhines1970edge,
  author = {Rhines, Peter B.},
  title = {Edge-, bottom-, and {Rossby} waves in a rotating stratified fluid},
  journal = {Geophysical Fluid Dynamics},
  volume = {1},
  number = {3--4},
  pages = {273--302},
  year = {1970},
  doi = {10.1080/03091927009365776},
}

@article{marques2014topographic,
  author = {Marques, Gustavo M. and Padman, Laurie and Springer, Scott R.
             and Howard, Susan L. and {\"O}zg{\"o}kmen, Tamay M.},
  title = {Topographic vorticity waves forced by {Antarctic} dense
             shelf water outflows},
  journal = {Geophysical Research Letters},
  volume = {41},
  number = {4},
  pages = {1247--1254},
  year = {2014},
  doi = {10.1002/2013GL059153},
}

@article{han2022trw,
  author = {Han, Xianxian and Stewart, Andrew L. and Chen, Dake
             and Lian, Tao and Liu, Xiaohui and Xie, Xiaohui},
  title = {Topographic {Rossby} Wave-Modulated Oscillations of
             Dense Overflows},
  journal = {Journal of Geophysical Research: Oceans},
  volume = {127},
  pages = {e2022JC018702},
  year = {2022},
  doi = {10.1029/2022JC018702},
}

@article{han2023controls,
  author = {Han, Xianxian and Stewart, Andrew L. and Chen, Dake
             and Liu, Xiaohui and Lian, Tao},
  title = {Controls of Topographic {Rossby} Wave Properties and
             Downslope Transport in Dense Overflows},
  journal = {Journal of Physical Oceanography},
  volume = {53},
  number = {7},
  pages = {1805--1820},
  year = {2023},
  doi = {10.1175/JPO-D-22-0237.1},
}

@article{guo2024seamount_trw,
  author = {Guo, Binbin and Shu, Yeqiang and Wang, Weiqiang and He, Gaowen and Liang, Qianyong and Zhang, Dongsheng and Yu, Lusha and Wang, Jun and Deng, Xiguang and Yang, Yong and Xie, Qiang and Deng, Yinan and Su, Danyi},
  title = {Observations of Intermittent Seamount-Trapped Waves and Topographic {Rossby} Waves around the Slope of a Low-Latitude Deep Seamount},
  journal = {Journal of Physical Oceanography},
  volume = {54},
  number = {1},
  pages = {281--297},
  year = {2024},
  doi = {10.1175/JPO-D-22-0121.1},
}

@article{wang2010,
  author = {Wang, Dong-Ping and Flagg, Charles N. and Donohue, Kathleen and Rossby, H. Thomas},
  title = {Wavenumber Spectrum in the {Gulf Stream} from Shipboard {ADCP} Observations and Comparison with Altimetry Measurements},
  journal = {Journal of Physical Oceanography},
  volume = {40},
  number = {4},
  pages = {840--844},
  year = {2010},
  doi = {10.1175/2009JPO4330.1},
}

@article{thompson2018antarctic,
  author = {Thompson, Andrew F. and Stewart, Andrew L. and Spence, Paul and Heywood, Karen J.},
  title = {The {Antarctic} Slope Current in a Changing Climate},
  journal = {Reviews of Geophysics},
  volume = {56},
  number = {4},
  pages = {741--770},
  year = {2018},
  doi = {10.1029/2018RG000624},
}

@article{GayenKlocker2024,
    author  = {Gayen, Bishakhdatta and Klocker, Andreas},
    title   = {Deep convection drives oceanic overturning},
    journal = {Physics Today},
    volume  = {77},
    number  = {6},
    pages   = {44--50},
    year    = {2024},
    doi     = {10.1063/pt.uvlg.jjol}
}

@article{ferrari2011processes,
  author  = {Ferrari, Raffaele and Ferreira, David},
  title   = {What processes drive the ocean heat transport?},
  journal = {Ocean Modelling},
  volume  = {38},
  number  = {3--4},
  pages   = {171--186},
  year    = {2011},
  doi     = {10.1016/j.ocemod.2011.02.013}
}

@article{whitehead2000circulation,
  title   = {Circulation and boundary layers in differentially heated rotating stratified fluid},
  author  = {Whitehead, J. A. and Pedlosky, J.},
  journal = {Dynamics of Atmospheres and Oceans},
  volume  = {31},
  number  = {1},
  pages   = {1--21},
  year    = {2000},
  issn    = {0377-0265},
  doi     = {10.1016/S0377-0265(99)00026-3}
}

@article{ong2024intrusions,
  author  = {Ong, Ellie Q. Y. and Doddridge, Edward and Constantinou, Navid C. and Hogg, Andrew McC. and England, Matthew H.},
  title   = {Intrinsically Episodic Antarctic Shelf Intrusions of Circumpolar Deep Water via Canyons},
  journal = {Journal of Physical Oceanography},
  year    = {2024},
  volume  = {54},
  number  = {5},
  pages   = {1195--1210},
  doi     = {10.1175/JPO-D-23-0067.1}
}

@article{stewart2012mixing,
  author  = {Stewart, Kial D. and Hughes, Graham O. and Griffiths, Ross W.},
  title   = {The Role of Turbulent Mixing in an Overturning Circulation Maintained by Surface Buoyancy Forcing},
  journal = {Journal of Physical Oceanography},
  year    = {2012},
  volume  = {42},
  number  = {11},
  pages   = {1907--1922},
  doi     = {10.1175/JPO-D-11-0242.1}
}

@article{Winters1995,
    author  = {Winters, Kraig B. and Lombard, Peter N. and
               Riley, James J. and D'Asaro, Eric A.},
    title   = {Available potential energy and mixing in
               density-stratified fluids},
    journal = {Journal of Fluid Mechanics},
    volume  = {289},
    pages   = {115--128},
    year    = {1995},
    doi     = {10.1017/S002211209500125X}
}

@article{Hughes2009,
    author  = {Hughes, Graham O. and Hogg, Andrew McC. and
               Griffiths, Ross W.},
    title   = {Available Potential Energy and Irreversible Mixing
               in the Meridional Overturning Circulation},
    journal = {Journal of Physical Oceanography},
    volume  = {39},
    number  = {12},
    pages   = {3130--3146},
    year    = {2009},
    doi     = {10.1175/2009JPO4162.1}
}

\end{document}